\documentclass[prb,showpacs,preprintnumbers,preprint,floatfix]{revtex4}
\usepackage{graphics}
\usepackage{graphicx}
\usepackage{dcolumn} 
\usepackage{bm}
\usepackage{float}
\usepackage{epsfig}
\begin{document}
\def\rhov{{\mbox{\boldmath{$\rho$}}}}
\def\tauv{{\mbox{\boldmath{$\tau$}}}}
\def\Deltav{{\mbox{\boldmath{$\Delta$}}}}
\def\Lambdav{{\mbox{\boldmath{$\Lambda$}}}}
\def\Thetav{{\mbox{\boldmath{$\Theta$}}}}
\def\Psiv{{\mbox{\boldmath{$\Psi$}}}}
\def\Phiv{{\mbox{\boldmath{$\Phi$}}}}
\def\sigmav{{\mbox{\boldmath{$\sigma$}}}}
\def\alphav{{\mbox{\boldmath{$\alpha$}}}}
\def\xiv{{\mbox{\boldmath{$\xi$}}}}
\def\oh{{\scriptsize 1 \over \scriptsize 2}}
\def\ot{{\scriptsize 1 \over \scriptsize 3}}
\def\of{{\scriptsize 1 \over \scriptsize 4}}
\def\tf{{\scriptsize 3 \over \scriptsize 4}}
\title{Phonons and their Coupling to Magnons in $n=2$
Ruddlesden-Popper Compounds.}

\author{A. B. Harris}

\affiliation{Department of Physics and Astronomy, University of
Pennsylvania, Philadelphia PA 19104}
\date{\today}
\begin{abstract}
The number of absorption and Raman optical modes for each Wyckoff orbit in the
high-temperature tetragonal (I4/mmm) parent lattice of the Ruddlesden-Popper
compounds Ca$_3$X$_2$O$_7$, with X=Mn or X=Ti is given. We analyze
the effect of sequential perturbations which lower the symmetry to Cmcm
and Cmc2$_1$ and finally include magnetic ordering. We determine
the power law behavior (within mean field theory) for the cross section
for photon absorption and Raman scattering of modes which appear as
the symmetry is successively lowered.  In the Cmc2$_1$ phase we
give a symmetry analysis to discuss the magnon-phonon coupling
which in other systems gives rise to ``electromagnons."
From our results we suggest several experiments to clarify the
phase diagram and other properties of these systems.
\end{abstract}
\maketitle

\section{INTRODUCTION}

The Ruddlesden-Popper (RP) compounds[\onlinecite{RP}] are those of the
type A$_{n+1}$B$_n$
C$_{3n+1}$, where the valences of the ions are A $=+2$, B $=+4$, and
C $=-2$ (usually C is oxygen).  At high temperature the crystal structure
for many of these compounds is tetragonal, but the structure can distort
in several ways as the temperature passes through  structural phase
transitions.  In the lowest symmetry phase, these systems are usually
ferroelectric and can also be magnetically ordered, although, in contrast
to many magnetoelectrics,[\onlinecite{ME1,ME2,ME3,ME4,ME5,ME6}] these
two degrees of freedom do not appear at the same temperature.
Nevertheless the RP systems do have many interesting couplings between
their magnetic and dielectric properties.  For example, their spontaneous
polarization can be rotated by the application of a magnetic field and
the weak ferromagnetic moment can be rotated by the application of an
electric field.[\onlinecite{CF,ABHCF}]  The symmetry of the RP's is
very similar to that of the Aurivillius compound SrBi$_2$Ta$_2$O$_9$
analyzed in detail by Perez-Mato {\it et al.}[\onlinecite{PEREZ}]
and many of our results should also apply to it.
Although there have been several first principles calculations of
the properties of CMO and/or CTO,[\onlinecite{CF,WHANG,CARDOSO}]
the calculations of CF are most relevant to the present paper.

\begin{figure}[h!]
\begin{center}
\caption{\label{FIG1} (Color online) The tetragonal phase of CMO or CTO
(space group I4/mmm).  The oxygen ions are the blue spheres, the Ca ions
are the open green circles, and the Mn or Ti ions (not shown) are at
the centers of the octahedra of oxygen ions. There are two molecules
per conventional unit cell.}
\vspace{0.2 in}
\includegraphics[width=5.0 cm]{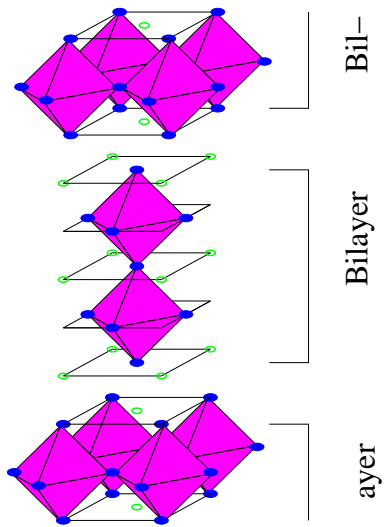}
\end{center}
\end{figure}

\begin{figure}[h!]
\begin{center}
\includegraphics[width=8.6 cm]{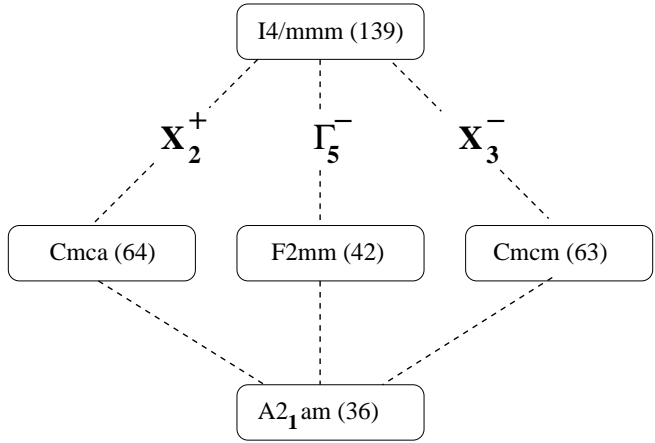}
\caption{\label{GSG} The group-subgroup structure of CMO and CTO.
We indicate the irreducible representations (irreps) by which the
tetragonal phase is distorted to reach one of the intermediate phases.
The numbering of the space groups follows Ref. \onlinecite{ITC}.
The superscripts on the irreps give the parity with respect to
spatial inversion.  The irrep $\Gamma_5^-$ is at zero wave vector and
the $X$ irreps are at the wave vectors ${\bf q}=(1/2,\pm 1/2,0)$.
In the final step to reach the A2$_1$am (equivalent to Cmc2$_1$) structure
the other two irreps are introduced simultaneously,[\onlinecite{ABHCF}]
so that Cmc2$_1$ can be regarded as the
result of simultaneous condensation of the three irreps.}  
\end{center}
\end{figure}

Here we will be interested in the two RP compounds Ca$_3$Mn$_2$O$_7$ (CMO)
and Ca$_3$Ti$_2$O$_7$ (CTO), which consist of bilayers of oxygen
octahedra enclosing a Mn or Ti ion.  It is believed that the high-temperature
phase is tetragonal (I4/mmm)[\onlinecite{TET,LAB}] 
and the low-temperature phase is
orthorhombic (Cmc2$_1$)[\onlinecite{ORTHO}] and in Fig.  \ref{GSG} we show
the three possible paths that connect these structures.[\onlinecite{CF}]
First principles calculations[\onlinecite{CF}] indicate that the most likely
scenario involves the appearance, as the temperature is lowered, of the
intermediate state Cmcm at a temperature we denote $T_>$. In fact,
the Cmcm phase has been observed in the isostructural compounds
LaCa$_2$Mn$_2$O$_7$[\onlinecite{GREEN}] at and below room temperature
and Bi$_{0.44}$Ca$_{2.56}$Mn$_2$O$_7$[\onlinecite{QIN}] at room teperature.
Accordingly, we will assume that this scenario applies to CMO and CTO although 
for these systems no intermediate phase has been observed up to now.
However, recent measurements on ceramic CMO find a clear pyroelectric signal
consistent with the onset of ferroelectric order close to
$T_>=280$K.[\onlinecite{LAWES}]  This transition is identified as the
temperature at which Cmc2$_1$ appears.  Since this ferroelectric transition
seems to be a continuous and well-developed one and since a direct continuous
transition between I4/mmm and Cmc2$_1$ is inconsistent with Landau 
theory,[\onlinecite{UTAH}] the seemingly inescapable conclusion is that
the phase for $T$ slightly greater than $T_>$ is {\it not} I4/mmm, but
is some phase which does not allow a spontaneous polarization.  Thus
the phase at temperature just above $T_>$ may be the Cmcm phase.
In this scenario the Cmc2$_1$ phase would appear at a lower temperature
$T_<$ (which experimental data implies is quite close to $T_>$).
The purpose of the present paper is to discuss the symmetry of
the zone center phonons and the Raman scattering cross section
as they are modified by the successive lowering of symmetry as the
temperature is reduced through the structural phase transitions at $T_>$ and
$T_<$.  Such a symmetry analysis can be done using
standard group theory methods.  Here we will supplement
that analysis with an analysis of the phonon cross sections from
which we determine their power law dependence on the order parameters
(OP's) which characterize the distortions from tetragonal symmetry to the
lower symmetry structures.  This enables us to
predict the temperature dependence of the newly allowed cross sections
which appear at the structural phase transitions. 
We then give a similar analysis of the magnon-phonon interaction.  As a
result of this coupling, the magnon absorption cross section (which usually
results from a magnetic dipole allowed transition) is enhanced by now being
electric-dipole allowed. This enhancement has led some
authors[\onlinecite{EM1,EM2,EM3,EM4,EM5,EM6}] to call such magnons
``electromagnons."

Briefly, this paper is organized as follows.  In Sec. II we give a symmetry
analysis of the zone center phonons in the tetragonal (I4/mmm) phase and
in Secs. III and IV we give similar analyses for the orthorhombic Cmcm
and Cmc2$_1$ phases.  We discuss the photon absorption and Raman scattering
of these modes. For modes which only appear as a consequence of a phase
transition we also determine the power-law dependence of their optical cross
sections on the emergent OP's.   In Sec. V we discuss the coupling of these
modes to magnetic excitations (magnons). Our results are summarized in Sec. VI.

\section{SYMMETRY ANALYSIS FOR THE TETRAGONAL STRUCTURE}

The occupied sites in space group I4/mmm for the $n=2$ RP system are given
in the first three columns of Table \ref{basis}, below.
The character table for the little group of ${\bf q}=0$
for the I4/mmm tetragonal structure is given in Table \ref{T4}.  For the one
dimensional irreducible representations (irreps) the characters are
the matrices.  For the two dimensional irreps and for the irreps for the 
$X$ points at ${\bf q}=(1/2,\pm 1/2,0)$ (in rlu's)
the matrices are given in Table \ref{IRREP}.[\onlinecite{FN1}]

%


\begin{small}
\begin{table}
\caption{\label{T4} Character table for the little group of ${\bf q}=0$
for the I4/mmm tetragonal structure. The operator numbers are those of Ref.
\onlinecite{MANDL}. The bottom row gives the dipole moment vector
${\bf P}$ or the Raman tensor ${\bf R}$, where $a$, $b$, $c$, etc. are
arbitrary constants.  For the two dimensional irreps
the two entries are the values of ${\bf P}$ and ${\bf R}$ for each of the
two degenerate states.  The entry is left blank for irreps which are optically
inactive.}
\vspace{0.2 in}
\begin{tabular} {|| c || c || c | c | c | c | c || c | c | c | c | c ||}
\hline \hline
Number & Operator &
\ \ $\Gamma_1^+$\ \ & \ \ $\Gamma_2^+$\ \ &\ \ $\Gamma_3^+$\ \ & \ \
$\Gamma_4^+$\ \ &\ \ $\Gamma_5^+$ \ \ &\ \ $\Gamma_1^-$\ \ &
\ \ $\Gamma_2^-$ \ \ &\ \ $\Gamma_3^-$\ \ &\ \ $\Gamma_4^-$\ \
&\ \ $\Gamma_5^-$ \ \ \\
\hline
$1$& $(xyz)$ & $1$ & $1$ & $1$ & $1$ & $2$ & $1$ & $1$ & $1$ & $1$ & $2$ \\
$37$,$40$ & $\ \ (yxz),(\overline y , \overline x, z)\ \ $
& $1$ & $-1$ & $-1$ & $1$ & $0$ & $-1$ & $1$ & $1$ & $-1$ & $0$ \\
$4$ & $(\overline x \overline yz)$
& $1$ & $1$ & $1$ & $1$ & $-2$ & $1$ & $1$ & $1$ & $1$ & $-2$ \\
$3,2$ & $(\overline xy \overline z), (x \overline y \overline z )$
& $1$ & $1$ & $-1$ & $-1$ & $0$ & $1$ & $1$ & $-1$ & $-1$ & $0$ \\
$39,38$ & $(\overline y x \overline z),(y \overline x \overline z)$
& $1$ & $-1$ & $1$ & $-1$ & $0$ & $-1$ & $1$ & $-1$ & $1$ & $0$ \\
\hline
$25$ & $(\overline x \overline y \overline z)$
& $1$ & $1$ & $1$ & $1$ & $2$ & $-1$ & $-1$ & $-1$ & $-1$ & $-2$ \\
$13,16$ & $(\overline y \overline x \overline z),(yx \overline z)$
& $1$ & $-1$ & $-1$ & $1$ & $0$ & $1$ & $-1$ & $-1$ & $1$ & $0$ \\
$28$ & $(xy \overline z)$
& $1$ & $1$ & $1$ & $1$ & $-2$ & $-1$ & $-1$ & $-1$ & $-1$ & $2$ \\
$27,26$ & $(x\overline yz),(\overline x y z)$
& $1$ & $1$ & $-1$ & $-1$ & $0$ & $-1$ & $-1$ & $1$ & $1$ & $0$  \\
$15,14$ & $(y\overline xz),(\overline y x z)$
& $1$ & $-1$ & $1$ & $-1$ & $0$ & $1$ & $-1$ & $1$ & $-1$ & $0$   \\
\hline
${\bf P}$ && $a(\hat i \hat i + \hat j \hat j)$ & $c(\hat i \hat i $ 
& & $d(\hat i \hat j $ & $e(\hat i \hat k + \hat k \hat i)$ &
& & $g \hat k$ & & $h \hat i$ \\
or ${\bf R}$ & & $+b \hat k \hat k$
& $-\hat j \hat j )$ & & $+\hat j \hat i)$
& $e(\hat j \hat k + \hat k \hat j)$ & & & & & $h \hat j$ \\
\hline \hline
\end{tabular}
\end{table}
\end{small}

\begin{table} [h!]
\caption{\label{IRREP} Representation matrices $M^{({\bf G})}({\cal O})$
for the operators ${\cal O}$ which are the generators of the irrep ${\bf G}$.
Here ${\bf r}'={\cal O}{\bf r}$ and the $\sigmav$'s are the Pauli matrices.
For the irreps ${\bf G}$ at zero wave vector the matrices for the
translations ${\bf T}_1 \equiv (x+1,y,z)$, ${\bf T}_2 \equiv (x,y+1,z)$,
and ${\bf T}_3\equiv (x+\frac{1}{2}, y+\frac{1}{2}, z- \frac{1}{2})$ are
unity. For the irreps at the ${\bf X}$ wave vectors the matrices are
${\bf M}({\bf T}_1)= {\bf M}({\bf T}_2)= -{\bf 1}$ and 
${\bf M}({\bf T}_3)=- \sigma_z$.  The first row and column for the ${\bf X}$
irreps refer to wave vector ${\bf q}_1$ These matrices are
related to those of Ref. \onlinecite{UTAH} by a unitary transformation
which, for zero wave vector takes $(\sigmav_x, \sigmav_y, \sigmav_z)$
into $(-\sigmav_x, -\sigmav_y, \sigmav_z)$ and for the wave vector ${\bf X}$
takes ($\sigmav_x, \sigmav_y, \sigmav_z)$
into $(-\sigmav_z, \sigmav_y, -\sigmav_x)$.}
\vspace{0.2 in}
\begin{tabular} {|| c | c c c ||}
\hline \hline
${\cal O}=$ & ${\cal R}_4$ & $m_d$ & $m_z$ \\
${\bf r}' = $ &\ \  $(\overline y, x,z)$\ \ &\ \ $(y,x,z)$\ \ &
\ \ $(x,y, \overline z)$\ \ \\
\hline
${\bf M}^{({\bf X}_1^+)}({\cal O})=$
& $- \sigmav_x$ & $1$ & ${\bf 1}$\\
${\bf M}^{({\bf X}_2^+)}({\cal O})=$
& $  \sigmav_x$ & $-1$ & ${\bf 1}$\\
${\bf M}^{({\bf X}_3^+)}({\cal O})=$
& $-i\sigmav_y$ & $\sigmav_z$ & $-{\bf 1}$\\
${\bf M}^{({\bf X}_4^+)}({\cal O})=$
& $i \sigmav_y$ & $-\sigmav_z$ & $-{\bf 1}$\\
\hline
${\bf M}^{({\bf X}_1^-)}({\cal O})=$
& $ - \sigmav_x$ & $-{\bf 1}$ & $-{\bf 1}$\\
${\bf M}^{({\bf X}_2^-)}({\cal O})=$
& $  \sigmav_x$ & ${\bf 1}$ & $-{\bf 1}$\\
${\bf M}^{({\bf X}_3^-)}({\cal O})=$
& $-i\sigmav_y$ & $-\sigmav_z$ &${\bf 1}$\\
${\bf M}^{({\bf X}_4^-)}({\cal O})=$
& $i \sigmav_y$ & $\sigmav_z$ &${\bf 1}$\\
\hline
${\bf M}^{(\Gamma_5^-)}({\cal O})=$ & $i \sigmav_y$ & $\sigmav_x$ & ${\bf 1}$ \\
${\bf M}^{(\Gamma_5^+)}({\cal O})=$ & $i \sigmav_y$ & $\sigmav_x$ & $-{\bf 1}$ \\
\hline \hline
\end{tabular}
\end{table}

\begin{table} [h!]
\caption{\label{basis} Basis functions $\Psi_1^{(X)}$ and $\Psi_2^{(X)}$ for
the distortion under irrep $X$,  where $X=5,3,2$ indicates irrep
$\Gamma_5^-$, $X_3^-$, and $X_2^+$, respectively.[\onlinecite{FN1}] For each
site we give the three components of the vector displacement.  We assume
normalization so that the sum of the squares of the components is unity.
The values of the structural parameters are given in Ref.
\onlinecite{TET}: $\rho=0.311$,
$\xi=0.100$, $\chi=0.205$, and $\tau=0.087$.  Since the a and b sites have
the same symmetry we will refer to them both as a sites.}
\vspace{0.2 in}
\begin{tabular} {|| c | c | c || c c c | c c c ||c c c | c c c ||c c c 
| c c c ||}
\hline \hline
Site &\ \  $n$\ \ & ${\bf r}$ & \multicolumn{3} {|c|} { $\Psi_{1,n}^{(5)}$} &
\multicolumn{3}{|c|} {$\Psi_{2,n}^{(5)}$} &
\multicolumn{3}{|c|} {$\Psi_{1,n}^{(3)}$} &
\multicolumn{3}{|c|} {$\Psi_{2,n}^{(3)}$} &
\multicolumn{3}{|c|} {$\Psi_{1,n}^{(2)}$} &
\multicolumn{3}{|c|} {$\Psi_{2,n}^{(2)}$} \\
\hline
\multicolumn{21} {|c||} {A sites} \\ \hline
e & 1& $(0,0,\rho +1/2) $ & $u$  & 0 & 0 & 0 & $u$  & 0 
& $a$ & $-a$ & 0 & $-a$ & $-a$ & 0 & 0 & 0 & 0 & 0 & 0 & 0 \\
e & 2& \ \ $(0,0,- \rho +1/2)$ \ \  &$u$&0&0\ \ &0&$u$&0\ \
&\ \ $a$ & $-a$ & 0 & $-a$ & $-a$ & 0 & 0 & 0 & 0 & 0 & 0 & 0 \\
b & 3& $(0,0,1/2)   $ & $v$  & 0 & 0 & 0 & $v$  & 0
& $b$ & $-b$ & 0 & $-b$ & $-b$ & 0 & 0 & 0 & 0 & 0 & 0 & 0 \\
\hline
\multicolumn{21} {|c||} {B sites} \\ \hline
e & 4& $(0,0,\xi)   $ & $w$  & 0 & 0 & 0 & $w$  & 0
& $c$ & $-c$ & 0 & $-c$ & $-c$ & 0 & 0 & 0 & 0 & 0 & 0 & 0 \\
e & 5& $(0,0,-\xi)  $ & $w$ & 0 & 0 & 0 &$w$  & 0
& $c$ & $-c$ & 0 & $-c$ & $-c$ & 0 & 0 & 0 & 0 & 0 & 0 & 0 \\
\hline
\multicolumn{21} {|c||} {O sites} \\ \hline
a & 6 & $(0,0,0)     $ & $x$  & 0 & 0 & 0 & $x$  & 0
& $d$ & $-d$ & 0 & $-d$ & $-d$ & 0 & 0 & 0 & 0 & 0 & 0 & 0 \\
e & 7 & $(0,0,\chi)  $ & $y$  & 0 & 0 & 0 & $y$  & 0
& $e$ & $-e$ & 0 & $-e$ & $-e$ & 0 & 0 & 0 & 0 & 0 & 0 & 0 \\
e & 8 & $(0,0,-\chi) $ & $y$ & 0 & 0 & 0 &$y$  & 0
& \ \ $e$ & $-e$ & 0 & $-e$ & $-e$ & 0 & 0 & 0 & 0 & 0 & 0 & 0 \\
g & 9 & $(0,1/2,\tau)$ & $z_1$ & 0 & 0 & 0 & $z_2$ &
$0$ & $0$ & $0$ & $f$ & $0$ & $0$ & $f$ & $-h$ & $-g$ & 0 & $-h$ & $g$ & 0 \\
g & 10 & $(0,1/2,-\tau)$ & $z_1$  &0 &0 & 0 & $z_2$ &0
& $0$ & $0$ & $-f$ & $0$ & $0$ & $-f$ & $-h$ & $-g$ & 0\ \
& $-h$ & $g$ & 0\ \ \\
g & 11 & $(1/2,0,\tau)$ & $z_2$  &0 &0 &0 & $z_1$ &0
& $0$ & $0$ & $-f$ & $0$ & $0$ & $f$ & $g$ & $h$ & 0 & $-g$ & $h$ & 0 \\
g & 12 & $(1/2,0,-\tau)$ & $z_2$  &0 &0 &0 & $z_1$ &0
& $0$ & $0$ & $f$ & $0$ & $0$ & $-f$ & $g$ & $h$ & 0 & $-g$ & $h$ & 0 \\
\hline \hline
\end{tabular}
\end{table}

We now discuss briefly the introduction of order parameters (OP's).  Starting
from the high-temperature tetragonal, I4/mmm, structure, distortions
of the irreps $\Gamma_5^-$, $X_3^-$ and $X_2^+$ are condensed as the
temperature is lowered
to eventually reach the low-symmetry low-temperature Cmc2$_1$ structure.
A tabulation of these symmetry-adapted distortion vectors, $\Psi_n^{(X)}$,
for irrep $X$ is given in Table \ref{basis}.  In the dominant component of the
distortion of irrep $X_2^+$ the octahedra are rotated about the crystal $c$
axis like interlocking gears.  The rotations in the upper ($z=\tau$) and lower
($z=-\tau$) planes of a bilayer are equal, so in the notation of Refs.
\onlinecite{ROTA} and \onlinecite{ROTB} these rotations are both denoted
$(0,0,\Theta)$. The dominant component of the 
distortion of irrep $X_3^-$ consists of 1) a rotation about the tetragonal
[1,0,0] with the rotation alternating in sign as one moves along the
[1,0,0] direction and 2) a rotation about the tetragonal [0,1,0] direction
with the rotation alternating in sign as one moves along the [0,1,0]
direction. Since this distortion is an even function of $z$, we identify 
this distortion as $(\Phi,\Phi,0)$ in the upper layer and 
$(-\Phi, -\Phi,0)$ in the lower layer. 

If we assume that the distortions are small, then the crystal
structure in any of the phases can be approximated as a distortion from the
tetragonal structure which is written as linear combination
of the distortions of the three irreps listed in Table \ref{basis}:
\begin{eqnarray}
\Phi &=& \sum_{m,X} Q_m^{(X)} \Psi_m^{(X)} \ , \nonumber
\end{eqnarray}
where $\Psi_m^{(X)}$ is the 12 component vector of irrep $X$ as in Table
\ref{basis}. Since the notation to distinguish the various irreps which we
will encounter is somehwat complicated, we summarize the notation in
Table \ref{NOTE}.
If we regard the basis vectors $\Psi_m^{(X)}$ as fixed, then application
of an operator ${\cal O}$ in the symmetry group of the tetragonal space
group to $\Phi$  will induce a transformation of the OP's as
[\onlinecite{ABHCF}]
\begin{eqnarray}
{\cal O} Q_m^{(X)} &=& \sum_k M^{(X)}_{k,m}({\cal O}) Q_k^{(X)} \ ,
\nonumber \end{eqnarray}
where the representation matrices ${\bf M}^{(X)} ( {\cal O} )$ for the
all the relevant operators ${\cal O}$ can be constructed from those given for
the generators of the irrep $X$ given in Table \ref{IRREP}. This equation
defines the symmetry of the OP's. 

\begin{table}
\caption{\label{NOTE} Notation for order parameters. First column: full
notation, subsequent column(s): abbreviated notation.}
\vspace{0.2 in}
\begin{tabular} {|c|c c c c c|}
\hline \hline
For & Full & & Abbrev & & Abrev' \\
\hline
$m=1,2$ & $Q_m^{(\Gamma_5^-)}$ & $\equiv$ & $Q_5^-(m)$ & $\equiv$ & $Q_5(m)$ \\ 
$m=1,2$ & $Q_m^{(\Gamma_5^+)}$ & $\equiv$ & $Q_5^+(m)$ & & \\ 
$p<5$ & $Q_1^{(\Gamma_p^-)}$ & $\equiv$ & $Q_p^-(0)$ & & \\
$p<5$ & $Q_1^{(\Gamma_p^+)}$ & $\equiv$ & $Q_p^+(0)$ & & \\
\hline
& $[Q_1^{(\Gamma_5^-)}+Q_2^{(\Gamma_5^-)}]\sqrt 2$ & $\equiv$ &  
$Z^-$ & $\equiv$ & $Z$ \\
& $[Q_1^{(\Gamma_5^-)}-Q_2^{(\Gamma_5^-)}]\sqrt 2$ & $\equiv$ &  
$Y^-$ & $\equiv$ & $Y$ \\
& $[Q_1^{(\Gamma_5^+)}+Q_2^{(\Gamma_5^+)}]/\sqrt 2$ & $\equiv$ &  
$Z^+$  & & \\
& $[Q_1^{(\Gamma_5^+)}-Q_2^{(\Gamma_5^+)}]/\sqrt 2$ & $\equiv$ &  
$Y^+$ & & \\
& $Q_1^{(\Gamma_3^-)}$ & $\equiv$ & $Q_3^-(0)$ & $\equiv$  &  $X$ \\
\hline
$m=1,2$ & $Q_m^{(X_2^+)}$ & $\equiv$ & $Q_2^+({\bf q}_m)$
& $\equiv$ & $Q_2({\bf q}_m)$ \\ 
$m=1,2$ & $Q_m^{(X_3^-)}$ & $\equiv$ & $Q_3^-({\bf q}_m)$ & $\equiv$ & 
$Q_3({\bf q}_m)$ \\ 
$m=1,2$, $p \not=2$ & $Q_m^{(X_p^+)}$ & $\equiv$ & $Q_p^+({\bf q}_m)$
& & \\ 
$m=1,2$, $p \not=3$ & $Q_m^{(X_3^-)}$ & $\equiv$ & $Q_3^-({\bf q}_m)$
& & \\ 
\hline \hline
\end{tabular}
\end{table}
As explained in Ref. \onlinecite{ABHCF} these distortions form
two families, one of which involves wave vector ${\bf q}_1 \equiv (1/2,1/2,0)$
and the other of which involves wave vector
${\bf q}_2 \equiv (1/2,-1/2,0)$.[\onlinecite{FAM}] The wave vectors
$(1/2,\pm 1/2,0)$ correspond to a doubling of the size of the primitive
unit cell.  The first family is described by the three OP's
$Q_3^-({\bf q}_1)$, $Q_2^+({\bf q}_1)$, and
$Z \propto Q_1^{(\Gamma_5^-)} +Q_2^{(\Gamma_5^-)}$, 
and the second by the OP's $Q_3^-({\bf q}_2)$, $Q_2^+({\bf q}_2)$, and
$Y\propto  Q_1^{(\Gamma_5^-)} - Q_2^{(\Gamma_5^-)}$.  In the absence of any
external fields, when the sample orders, one expects an equal population of
domains of the two families.  However, it is possible to select a
single family by invoking the coupling[\onlinecite{ABHCF}]
\begin{eqnarray}
V &=& a \epsilon_{xy} [Q_k({\bf q}_1)^2  - Q_k({\bf q}_2)^2] \ ,
\label{STRESSEQ} \end{eqnarray}
where $\epsilon_{xy}$ is a shear component of strain, and $k=2$ or 3.
One can utilize this interaction by cooling the sample through
$T_>$ in the presences of a shear stress conjugate to $\epsilon_{xy}$.
This will preferentially select ${\bf q}_1$ or ${\bf q}_2$, depending on
the sign of the stress and the sign of the constant $a$. It is
possible that the selection of the wave vector can also be accomplished
by cooling in an electric or magnetic field.[\onlinecite{ABHCF}]

Our aim is describe the evolution of the cross section for a) zero wave
vector photon absorption and b) Raman scattering in terms of the various OP's.

The single photon absorption process results from the electromagnetic
coupling $V_{\rm E} = - {\bf p} \cdot {\bf E}$, where ${\bf E}$ is the
electric field of the incident electromagnetic wave and ${\bf p}$ is
the dipole moment operator of the system.
The symmetry indicated in Tables \ref{T4} and \ref{IRREP} indicates that in
the tetragonal phase one has
\begin{eqnarray}
p_x & = & a_x Q_5(1) \ , \ \ 
p_y = a_y Q_5(2) \ , \ \ 
p_z = a_z Q_3^-(0) \ ,
\nonumber \end{eqnarray}
with $a_y=a_x$.  Note: in reality we ought to index
the modes by an additional index to distinguish between different
modes of the same symmetry.  Thus, for instance,
\begin{eqnarray}
p_x & = & \sum_{k=1}^{n(\Gamma_5^-)} a_{k,x} Q_5(1)_k \ ,
\label{EQPX} \end{eqnarray}
where $n(\Gamma_5^-)$ is the number of occurrences of $\Gamma_5^-$,
which, in this case, is [see Eq. (\ref{REP1EQ}), below]  $n(\Gamma_5^-)=7$.
For simplicity we will usually not display this index explicitly, but in
all our results sums over such an index (or indices) are implied.  We discuss 
the use of Eq. (\ref{EQPX}) in Appendix \ref{APPA}.  The
absorption cross section is proportional to the square of the scalar product
of the polarization vector of the incident photon and the dipole moment
vector for the irrep in question.[\onlinecite{FERRARO, RAMAN}] (For the two
dimensional irrep one has to sum the squares of the two scalar products.)

Likewise, the Raman scattering process results from the electromagnetic
coupling
$V_{\rm E} = - \sum_{\beta \gamma} \alpha_{\beta \gamma} E_\beta E_\gamma$,
where $\alphav$ is the polarizability tensor, which depends on the
phonon displacements as
\begin{eqnarray}
\alpha_{\beta \gamma} &=& \alpha_{\beta \gamma}^{(0)} +
\sum_\tau \frac{\partial \alpha_{\beta \gamma}} {\partial Q_\tau} Q_\tau \ .
\nonumber \end{eqnarray}
We consider the cross section (or intensity) for a process in which the 
incident photon has an electric field vector in the $\beta$ direction and
the scattered photon has an electric field vector in the $\gamma$
direction (we call this the $\beta$-$\gamma$ polarization) and a
phonon $Q_\tau$ is created (or destroyed).  The cross section for this
process is proportional to the square of the tensor product of the Raman
tensor $\partial \alpha_{\beta \gamma} / \partial Q_\tau$
with the polarizations of the incoming and outgoing 
photons.[\onlinecite{FERRARO,RAMAN}]
(For the two dimensional irreps one sums these squares over the two states
of the irreps.) For both absorption and scattering we implicitly assume
that the photon wave length is very large in comparison to the size of
the unit cell.

Now we determine which modes (at zero wave vector) are allowed in the
tetragonal lattice.  Using standard methods
we find that the reducible representations $\Gamma$ corresponding to the
space of zero wave vector displacements of the Wyckoff orbits a, e, and g,
are decomposed into their irreps as
\begin{eqnarray}
\Gamma (a)  &=& \Gamma_5^- + \Gamma_3^- \ , \nonumber \\ 
\Gamma (e)  &=& \Gamma_5^- + \Gamma_3^- + \Gamma_5^+ + \Gamma_1^+ \ , 
\nonumber \\ 
\Gamma(g)  &=& 2 \Gamma_5^- + \Gamma_4^- + \Gamma_3^- + 2 \Gamma_5^+ 
+ \Gamma_2^+ + \Gamma_1^+  \ , 
\label{TETIRREP} \end{eqnarray}
so that the reducible representation $\Gamma$ corresponding to the space
of the vector displacements of all 12 ions in the primitive unit cell
has the decomposition into irreps as
\begin{eqnarray}
\Gamma &=& 2 \Gamma (a) + 3 \Gamma (e) + \Gamma(g) =
7 \Gamma_5^- + \Gamma_4^- \nonumber \\ &&
+ 6 \Gamma_3^- 
+ 5 \Gamma_5^+ + 4 \Gamma_1^+ + \Gamma_2^+ \ .
\label{REP1EQ} \end{eqnarray}
The allowed absorption modes [$x$ and $y$ from $\Gamma_5^-(0)$ and 
$z$ from $\Gamma_3^-(0)$] are listed in Table \ref{T9a}.  Thus we find
that in the tetragonal phase there are 20 absorption modes, 6 from
$\Gamma_3^-$ which have $z$ polarization and 7 pairs of doubly
degenerate modes (one with $x$ polarization and one with $y$ polarization)
coming from $\Gamma_5^-$. Note that of these 20 absorption modes three are
acoustic (zero frequency) and are not observed in IR experiments.
There are also 15 Raman active modes, five nondegenerate modes coming from
irreps $\Gamma_1^+$ and $\Gamma_2^+$ and 5 pairs of doubly degenerate
modes coming from $\Gamma_5^+$.  Because the lowest symmetry sites of the
space group I4/mmm are not occupied in CMO or CTO, it happens that there
are no modes for which the Raman polarization $xy$ is nonzero.
Indeed, this result can be used as a signature of the tetragonal phase
of CTO or CMO.

%
%
%

\begin{table} [h!]
\caption{\label{T9a} Number of modes $Q$ of the tetragonal phase with their
polarizations $x$, $y$, or $z$, for Wyckoff orbits, RP$(n)$ for $n=a,b,c$.  The 
phonon modes $Q_5^\pm (1)$ and $Q_5^\pm (2)$ are degenerate in energy.
To identify with modes in the orthorhombic phase $z_t \rightarrow X_O$,
$(x_t,y_t)\rightarrow (Y_O,Z_O)$, where the subscript $O$ ($t$) refers to 
the orthorhombic (tetragonal) coordinate system.  Pol. denotes polarization
in a notation where $xy$ indicates that the Raman tensor is 
${\bf R} = {\rm const} (\hat i \hat j + \hat j \hat i)$,
$X_OY_O$ indicates that ${\bf R}={\rm const} (\hat I \hat J + \hat J \hat I)$,
where lower case unit vectors are tetragonal and capitals are orthorhombic,
and similarly for other polarizations.  Also $Y_O^2-Z_O^2$ indicates that
${\bf R}= {\rm const.} (\hat J \hat J - \hat K \hat K)$ and 
$X_O^2$, $(Y_O^2+Z_O^2)$ indicates that ${\bf R} = a\hat I \hat I
+b(\hat J \hat J + \hat K \hat K)$, where $a$ and $b$ are constants.
(The orthorhombic polarizations are those when
the tetragonal symmetry is broken.) The row labeled TOT gives the
total number of modes for 2 a orbits, 3 e orbits, and 1 g orbit.}

\vspace{0.2 in}
\begin{tabular} {||c || c | c | c || c | c | c | c | c ||}
\hline \hline
& \multicolumn{3} {c||} {Absorption} & \multicolumn{5} {c||} 
{Raman Scattering}
\\
\hline
Mode$_t$ &\ $Q_5^- (1)$\ &\ $Q_5^- (2)$\ &\  $Q_3^-(0)$\
& \ $Q_5^+(1)$\ & \ $Q_5^+(2)$\  & \ $Q_4^+(0)$\ &
 \ $Q_2^+(0)$ \ & \ $Q_1^+ (0)$ \ \\
\ \ Pol.$_t$\ \ & $x$ & $y$ & $z$ & $zx$ & $zy$ & $xy$ & $x^2-y^2$ &
$x^2+y^2,z^2$ \\
\hline
Mode$_O$ & $Y$ & $Z$ & $X$ & $Y^+$ & $Z^+$ & & & \\
Pol.$_O$ & $Y$ & $Z$ & $X$ & $XY$ & $XZ$ &
$(Y^2-Z^2)$ & $YZ$ & $X^2,(Y^2+Z^2)$ \\
\hline \hline
RP(a) & 1 & 1 & 1 & 0 & 0 & 0 & 0 & 0 \\
RP(e) & 1 & 1 & 1 & 1 & 1 & 0 & 0 & 1 \\
RP(g) & 2 & 2 & 1 & 2 & 2 & 0 & 1 & 1 \\
\hline \hline
TOT$^{\rm a}$ & 7 & 7 & 6 & 5 & 5 & 0 & 1 & 4 \\
\hline \hline
\end{tabular}

\vspace{0.2 in} \noindent
a)  Here and below the absorption modes include an acoustic
mode for each polarization.

\end{table}

\section{Orthorhombic Space Group Cmcm (\# 63)}

\subsection{Analysis based on Cmcm symmetry}

\begin{figure}[h!]
\begin{center}
\includegraphics[width=5.0 cm]{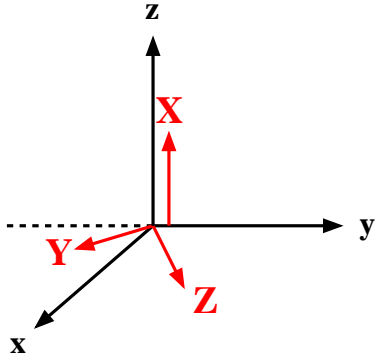}
\caption{\label{TO} (Color online) Tetragonal (large black arrows) and
orthorhombic (small red arrows) axes.}
\end{center}
\end{figure}

We now consider the modifications in the optical spectrum when we condense
irrep $X_3^-$ at wave vector ${\bf q}_1 = (1/2,1/2,0)$ to go into the space
group Cmcm.  The transformation from tetragonal to orthorhombic coordinates
is illustrated in Fig. \ref{TO} and is given by
\begin{eqnarray}
X_O &=& z_t , \hspace{0.4 in}
Y_O = \frac{x_t-y_t}{2} - \frac{1}{4} , \hspace{0.4 in}
\nonumber \\ Z_O &=& \frac{x_t+y_t}{2} - \frac{1}{4}  ,
\label{TRANSEQ} \end{eqnarray}
where here and below we use lower case letters for tetragonal coordinates and
capitals for orthorhombic coordinates.  The O unit cell is twice as large as
the t unit cell.  That is why $x_t \pm y_t$ is divided by 2 rather than by
$\sqrt 2$.  In Table \ref{T7} we give the character table for the little group 
of ${\bf q}=0$ for Cmcm as well as the characters for the reducible
representation associated with the vector displacement over each of the
Wyckoff orbits.

\begin{table}
\caption{\label{T7} Character table for the little group of ${\bf q}=0$ for
Cmcm and (in the last row) how they transform. The last three columns
give the character vectors for displacements for the Wyckoff orbits of the
RP lattice.  The $\Gamma_1^+$ modes are Raman active with Raman tensor
${\bf R} = a \hat I \hat I + b \hat J \hat J + c \hat K \hat K$.
All coordinates are orthorhombic. The modes of irrep $\Gamma_3^-$ transform
like $XYZ$ and are silent.}
\vspace{0.2 in}
\begin{tabular} {|| c || c || c | c | c | c || c | c | c | c ||c|c|c||}
\hline \hline
Class & Operator &
\ \ $\Gamma_1^+$\ \ & \ \ $\Gamma_2^+$\ \ &\ \ $\Gamma_3^+$\ \ & \ \
$\Gamma_4^+$\ \ &\ \ $\Gamma_1^-$\ \ & \ \ $\Gamma_2^-$ \ \ 
&\ \ $\Gamma_3^-$\ \ &\ \ $\Gamma_4^-$\ \ \ \ & \ \ a \ \ & \ \ e \ \ & \ \
g \ \\
\hline
$E$& $(X,Y,Z)$ & $1$ & $1$ & $1$ & $1$ & $1$ & $1$ & $1$ & $1$ 
& 6 & 12 & 24 \\
$2_X$ & $\ \ (X,\overline Y, \overline Z)\ \ $
& $1$ & $-1$ & $1$ & $-1$ & $1$ & $-1$ & $1$ & $-1$ & 0 & 0 & $-8$ \\
$m_Z$ & $(X,Y, \frac{1}{2} -Z)$
& $1$ & $-1$ & $-1$ & $1$ & $1$ & $-1$ & $-1$ & $1$ & 2 & 4 & 0 \\
$m_Y$ & $(X, \overline Y, \frac{1}{2} + Z)$
& $1$ & $1$ & $-1$ & $-1$ & $1$ & $1$ & $-1$ & $-1$ & 0 & 0 & 0 \\
\hline
${\cal I}$ & $(\overline X, \overline Y, \overline Z)$
& $1$ & $1$ & $1$ & $1$ & $-1$ & $-1$ & $-1$ & $-1$ & 0 & 0 & 0 \\
$m_X$ & $(\overline X, Y,  Z)$
& $1$ & $-1$ & $1$ & $-1$ & $-1$ & $1$ & $-1$ & $1$ & 2 & 0 & 0 \\
$2_Z$ & $(\overline X, \overline Y, \frac{1}{2} + Z)$
& $1$ & $-1$ & $-1$ & $1$ & $-1$ & $1$ & $1$ & $-1$ & 0 & 0 & 0 \\
$2_Y$ & $(\overline X, Y, \frac{1}{2} - Z)$
& $1$ & $1$ & $-1$ & $-1$ & $-1$ & $-1$ & $1$ & $1$ & $-2$ & 0 & 0 \\
\hline \hline
\multicolumn{2} {|c|} {${\bf P}$ or ${\bf R}$} & 1 & $XZ$ & $YZ$ & $XY$ 
& $X$ & $Z$ & $XYZ$ & $Y$ & & & \\
\hline \hline
\end{tabular}
\end{table}

To find out how many Cmcm optically active modes there are, we need to
convert the tetragonal coordinates into the orthorhombic coordinates but
we do not need to explicitly include the distortion that makes the
orthorhombic structure.
From the parent tetragonal orbit a we have the orthorhombic orbit
\begin{eqnarray}
{\bf r}_1 &=& (0,0,0)_t = (0, - \frac{1}{4} , - \frac{1}{4})_o \ , \nonumber \\
{\bf r}_2 &=& (1,0,0)_t = (0, \frac{1}{4}, \frac{1}{4})_o \ , \ \
\nonumber \end{eqnarray}
where the subscript t means tetragonal and the subscript o means Cmcm.
We find that
\begin{eqnarray}
\Gamma (a_{Cmcm}) &=& \Gamma_1^+ + \Gamma_3^+ + \Gamma_4^+ 
+ \Gamma_1^- + \Gamma_2^- + \Gamma_4^- \ .
\label{AEQ} \end{eqnarray}

For the parent tetragonal orbit e we have
\begin{eqnarray}
{\bf r}_1 &=& (0,0,z)_t = (z, - \frac{1}{4}, - \frac{1}{4})_o \ , \nonumber \\
{\bf r}_2 &=& (1,0,z)_t = (z, \frac{1}{4}, \frac{1}{4})_o \ , \nonumber \\
{\bf r}_3 &=& (0,0,-z)_t = (-z, - \frac{1}{4}, -\frac{1}{4})_o \ , \nonumber \\
{\bf r}_4 &=& (1,0,-z)_t = (-z, \frac{1}{4}, \frac{1}{4})_o \ , \nonumber
\end{eqnarray}
and we find that
\begin{eqnarray}
\Gamma (e_{Cmcm}) &=& 2 \Gamma_1^+ + \Gamma_2^+ + \Gamma_3^+ + 2\Gamma_4^+ 
+ 2\Gamma_1^- + \Gamma_2^- \nonumber \\ && \ + \Gamma_3^- + 2 \Gamma_4^- \ .
\label{EEQ} \end{eqnarray}

Similarly, for the parent tetragonal orbit g we find that
\begin{eqnarray}
\Gamma (g_{Cmcm}) &=& 2 \Gamma_1^+ + 4\Gamma_2^+ + 2\Gamma_3^+ + 4\Gamma_4^+ 
+ 2\Gamma_1^- \nonumber \\ && \ + 4\Gamma_2^- + 2\Gamma_3^- + 4 \Gamma_4^- \ .
\label{GEQ} \end{eqnarray}
Thus the decomposition of the reducible representation for the displacements of
all ions in the unit cell is
\begin{eqnarray}
\Gamma &=& 2 \Gamma(a_{Cmcm}) + 3 \Gamma(e_{Cmcm}) + \Gamma(g_{Cmcm}) 
\nonumber \\ &=& 10 \Gamma_1^+ +7 \Gamma_2^+ +7 \Gamma_3^+ +12 \Gamma_4^+
+10 \Gamma_1^- \nonumber \\ && \ 
+ 9  \Gamma_2^- +5 \Gamma_3^- +12 \Gamma_4^- \ . \nonumber
\end{eqnarray}
So in the Cmcm phase we have 10 modes with $X_O$ polarization, 12 modes with
$Y_O$ polarization, and 9 modes with $Z_O$ polarization, as given in Table
\ref{T9}. (The degeneracies of the tetragonal structure are removed in the
Cmcm structure.) Results for the Raman-active modes are also given in 
Table \ref{T9}.

\begin{table} [h!]
\caption{\label{T9} As Table \ref{T9a}.  Modes $Q$ in the orthorhombic Cmcm
phase for the Wyckoff orbits descended from the tetragonal orbits.  The irreps 
$\Gamma_n^{\pm}$ are those of Table \ref{T7} for the Cmcm phase. 
In the next to last row we give the total number of modes
for 2 a orbits, 3 e orbits and 1 g orbit.  In the last row we
repeat the results for the total number of modes for the tetragonal system.}
\vspace{0.2 in}
\begin{tabular} {||c || c | c | c || c | c | c | c ||}
\hline \hline
& \multicolumn{3} {c||} {Absorption} & \multicolumn{4} {c||} {Raman} \\
\hline
IRREP &\ \ $\Gamma_1^-$ \ \ &\ \ $\Gamma_2^- $\ \ &\ \  $\Gamma_4^-$\ \
& \ \ $\Gamma_1^+$\ \ &\ \ $\Gamma_2^+$\ \
&\ \ $\Gamma_3^+$\ \ &\ \ $\Gamma_4^+$\ \ \\
\ \ Polarization\ \ & $X_O$ & $Z_O$ & $Y_O$ \ \
& $X_O^2$, $Y_O^2,Z_O^2$ & $X_OZ_O$ & $Y_OZ_O$ & $X_OY_O$ \\
\hline \hline
RP(a) & 1 & 1 & 1 & 1 & 0 & 1 & 1 \\
RP(e) & 2 & 1 & 2 & 2 & 1 & 1 & 2 \\
RP(g) & 2 & 4 & 4 & 2 & 4 & 2 & 4 \\
\hline \hline
TOT & 10 & 9 & 12 & 10 & 7 & 7 & 12 \\
\hline
TET & 6 & 7 & 7 & 4 & 5 & 1 & 5 \\
\hline \hline
\end{tabular}
\end{table}

\subsection{Analysis based on lowering of tetragonal symmetry}

We now discuss the phonon modes by analyzing the perturbative effect of
distortions from the tetragonal parent structure.
Compared to the previous analysis, this analysis provides information
on how the intensity of the new Cmcm modes depends on the amplitude of
the orthorhombic distortion from tetragonal symmetry.
Let us suppose that $Q_3^-({\bf q}_1)$ has condensed due to the phase
transition at temperature $T=T_>$ into the Cmcm phase.  (Near this phase
transition one can choose modes so that only a single mode of this symmetry
is involved.  As the temperature is lowered below $T_>$ other modes of this
symmetry will progressively be introduced.) To see how the existence of
the new order parameter can modify modes of the tetragonal system
consider the Hamiltonian  ${\cal H}_5$ for $Q_{5,n}^-$,
\begin{eqnarray}
{\cal H}_5 &=& \frac{1}{2} \kappa_{5} \Biggl( Q_5(1)^2 + Q_5(2)^2
\Biggr) + \frac{1}{2} \kappa_{3-} [Q_{3}^-({\bf q}_1)]^2 \nonumber \\ && \
+ \sum_n b_n Q_5(1) Q_5(2) \langle Q_3^-({\bf q}_1) \rangle^n \ ,
\label{EQHAM} \end{eqnarray}
where the $\kappa$'s are stiffnesses and we put angled brackets around
$Q_3({\bf q}_1)$ to indicate that these higher than quadratic terms give rise
to effective quadratic terms written here when $Q_3^-({\bf q}_1)$ (whose
value minimizes the free energy) is regarded as the constant
$\langle Q_3^- ({\bf q}_1)\rangle$. To make the
above Hamiltonian invariant under all the operations of the parent 
tetragonal system, we would have to include analogous terms involving
${\cal R}_4 Q_3^-({\bf q}_1) = Q_3^-({\bf q}_2)$, where
${\bf q}_2 = (1/2,-1/2,0)$.  However, since we are considering a scenario
in which $Q_3^-({\bf q}_2)$ is zero,
we do not need to write these terms here.
Since $b_1=0$ due to wave vector conservation, Eq. (\ref{EQHAM}) is
\begin{eqnarray}
{\cal H}_5 &=& \frac{1}{2} \Biggl( \kappa_5 - b_2
\langle Q_3^-({\bf q}_1) \rangle^2 \Biggr) \frac{[Q_5(1) -
Q_5(2)]^2} {2} \nonumber \\ && \ + \frac{1}{2} \Biggl( \kappa_5 + b_2
\langle Q_3^-({\bf q}_1) \rangle^2 \Biggr) \frac{[Q_5(1) +
Q_5(2)]^2} {2}  \nonumber \\ & \equiv &
\frac{1}{2}\kappa_Y Y^2 + \frac{1}{2}\kappa_Z Z^2 \ ,
\label{H5EQ}\end{eqnarray}
where $Y= [Q_5(1) - Q_5(2)]/ \sqrt 2$, and
$Z = [Q_5(1) + Q_5(2)]/ \sqrt 2$.  This notation (of Table \ref{NOTE})
is motivated by the labelling of the axes in the
orthorhombic phase (see Eq. (\ref{TRANSEQ}) and Fig. \ref{TO}).
We therefore see that the presence of a nonzero value of 
$\langle Q_3^-({\bf q}_1)\rangle$ lifts the degeneracy in energy
between the two $Q_5^-$ modes and leads to the proper
linear combinations $Y$ and $Z$ as
the new normal modes of the Cmcm structure. The 
analogous splitting occurs in the same way for the $Q_5^+$ modes
and leads to the redefinition of normal modes as
$Y^+ = [Q_5^+(1) - Q_5^+(2)]/\sqrt 2$ and
$Z^+ = [Q_5^+(1) + Q_5^+(2)]/\sqrt 2$.

\subsubsection{Induced Absorption Modes}

We next investigate the mixing of modes due to the nonzero value of
$\langle Q_3^-({\bf q}_1)\rangle$.  The Hamiltonian we consider
for the mixing of absorptive modes is ${\cal H}_3 + {\cal H}_5 +V_5$, where
\begin{eqnarray}
{\cal H}_3 &=& \frac{1}{2} \kappa_3 \sum_k Q_3^- ({\bf q}_k)^2 \ ,
\nonumber \end{eqnarray}
\begin{eqnarray}
V_5 &=&  \frac{1}{2} \sum_n \Biggl( \kappa_{U_n} U_n^2
+ \kappa_{V_n} V_n^2 + \kappa_{W_n} W_n^2 \Biggr) \nonumber \\
&& \ + \sum_n \langle Q_3^-({\bf q}_1) \rangle^n
\Biggl( a^{(n)} U_n Z + b^{(n)} V_n Y 
\nonumber \\ && \ + c^{(n)} W_n Q_3^-(0) \Biggr) \ ,
\label{VVEQ} \end{eqnarray}
where $U_n$, $V_n$, $W_n$ are single phonon operators (to be determined) which,
as we will see, become optically active by virtue of their coupling to
the optically active $Z$, $Y$, or $Q_3^-(0)$ modes by the interaction
$V_5$ at order $\langle Q_3^-({\bf q}_1) \rangle^n$. To identify these
newly active modes, we analyze the symmetry of the coupling terms in
Eq. (\ref{VVEQ}).  Note that wave vector conservation indicates that
terms in Eq. (\ref{VVEQ}) with $n$ even involve newly active modes at zero
wave vector and those with $n$ odd involve newly active zero wave vector
modes which, before the distortion, were at wave vector ${\bf q}_1$ in the
tetragonal Brillouin zone.  The character table for the ${\bf X}$ wave vectors
was given in Table \ref{IRREP}.

To see more specifically what the Hamiltonian of Eqs. (\ref{H5EQ}) and
(\ref{VVEQ}) implies, consider only the terms involving $U_n$ and
$Z$:
\begin{eqnarray}
{\cal H}_X &=& \frac{1}{2}\kappa_Z Z^2 + \frac{1}{2}
\sum_n \kappa_{U_n} U_n^2 \nonumber \\ && \  +  \sum_n
a^{(n)} \langle Q_3^-({\bf q}_1) \rangle^n U_n Z \ .
\label{EQ31} \end{eqnarray}
In carrying out perturbation theory relative to the tetragonal phase
we use the correct linear combinations $Z$ and $Y$.  At the
moment we ignore the
perturbation in the energy due to this interaction which was addressed by
Eq. (\ref{H5EQ}).  Our interest here lies in the mixing of modes. When the
quadratic Hamiltonian of Eq. (\ref{EQ31}) 
is diagonalized, the new modes ${\tilde U}_n$ perturbed
through their coupling to $Z$ will now be
\begin{eqnarray}
\tilde U_n &=& U_n - \frac{ a^{(n)} \langle Q_3^-({\bf q}_1)\rangle^n
Z} {\kappa_{U_n} - \kappa_Z} \ .
\label{MIX} \end{eqnarray}
By analogy we will also have new modes
\begin{eqnarray}
\tilde V_n &=& V_n - \frac{ b^{(n)} \langle Q_3^-({\bf q}_1)\rangle^n
Y} {\kappa_{V_n} - \kappa_Y} \nonumber
\end{eqnarray}
\begin{eqnarray}
\tilde W_n &=& W_n - \frac{ c^{(n)} \langle Q_3^-({\bf q}_1)\rangle^n
Q_3^-(0)} {\kappa_{W_n} - \kappa_{3}} \ . \nonumber 
\end{eqnarray}
What this means is that the modes $U_n$, $V_n$, and $W_n$
(which we will identify in a moment) are optically active because they
are linear combinations with the optically active modes $Z_k$, $Y_k$,
and $Q_3(0)_k$, respectively, whose polarizations they inherit.
For quantitative work (see Appendix A), it is necessary to keep
track of the coupling to the different optically active modes,
$Z_k$, $Y_k$, and $Q_3^-(0)_k$.  For simplicity, since we are mainly
interested in which modes are newly induced, we need not keep the sum
over $k$.

Near $T_>$, $\langle Q_3({\bf q}_1) \rangle$ will be proportional to
$(T_>-T)^\beta$, where $\beta=1/2$ in mean field theory.  Thus the temperature
dependence of terms with different powers of this variable will be 
distinguishably different.  Analogous terms for the scenario when the
wave vector ${\bf q}_2$ has condensed can be obtained by applying the
operator ${\cal R}_4$ to ${\cal H}_3 + {\cal H}_5+V_5$.  Now we 
study how symmetry restricts the $U$'s, $V$'s and $W$'s of Eq. (\ref{VVEQ}).
Since ${\cal R}_4$ is not a relevant symmetry operation, we require
${\cal H}_3 + {\cal H}_5 +V_5$ to be invariant under ${\cal I}$, $m_d$, and
$m_z$.  If $n$ is odd (even), then $U_n$, $V_n$, and $W_n$ are at wave
vector ${\bf q}_1$ (wave vector zero).  To be invariant under inversion,
$U_n$, $V_n$, and $W_n$ have to have parity $(-1)^{n+1}$ under inversion.
Also $U_n$ and $V_n$ must be even under $m_z \equiv (xy\overline z)$
and $W_n$ must be odd under $(xy\overline z)$.  Finally we have to consider
the effect of $m_d=(yxz)$.  From Tables \ref{T4} and \ref{IRREP} we find that
\begin{eqnarray}
m_d Y &=& - Y \ , \ \
m_d Z = Z \ , \nonumber \\
m_d Q_3^-(0) &=& Q_3^-(0) \ , \ \
m_d Q_3^-({\bf q}_1) = - Q_3^-({\bf q}_1) \ . \nonumber
\end{eqnarray}
For $V_5$ to be invariant under $m_d$, we therefore must have
\begin{eqnarray}
m_d U_n & = & (-)^{n} U_n \ , \ \ \ \
m_d V_n  =  (-)^{n+1} V_n \ , \nonumber \\
m_d W_n  &=&  (-)^{n} W_n \ .
\label{EQMDV} \end{eqnarray}

Note that all the above constraints depend on $n$ only through $(-1)^n$,
so that to find the leading temperature dependences just below $T_>$,
we only need to consider two distinct cases, $n=1$ and $n=2$.
First consider $n=1$.  Then
\begin{eqnarray}
{\cal I} U_1 ({\bf q}_1) &=& U_1({\bf q}_1) \ , \ \ \ \
m_z U_1({\bf q}_1) = U_1({\bf q}_1) \ , \nonumber \\
m_d U_1({\bf q}_1) &=& -U_1({\bf q}_1) \ , \ \ \ \
{\cal I} V_1({\bf q}_1)  = V_1({\bf q}_1) \ , \nonumber \\
m_z V_1({\bf q}_1) &=& V_1({\bf q}_1) \ , \ \ \ \
m_d V_1({\bf q}_1) = V_1({\bf q}_1) \ , \nonumber \\
{\cal I} W_1 ({\bf q}_1) &=& W_1({\bf q}_1) \ , \ \ \ \
m_z W_1({\bf q}_1) = - W_1({\bf q}_1) \ , \nonumber \\
m_d W_1({\bf q}_1) &=& - W_1({\bf q}_1) \ , \nonumber
\end{eqnarray}
so that
\begin{eqnarray}
U_1 &=& \alpha_1Q_2^+({\bf q}_1) \ , \ \
V_1 = \beta_1 Q_1^+({\bf q}_1) \ , \ \ 
W_1 = \gamma_1 Q_4^+({\bf q}_1) \  , \nonumber
\end{eqnarray}
where $\alpha_1$, $\beta_1$, and $\gamma_1$ are constants whose values are 
not fixed by symmetry.  The meaning of $U_1$ is that because of the
allowed coupling between $Z$ and {\it any} operator having the symmetry
of $Q_2^+({\bf q}_1)$ this coupling will render all such modes
optically active with polarization $Z$ inherited from coupling to $Z$.
This coupling can be analoyzed for each Wykoff orbit, as we do
in Table \ref{T14}. To summarize: the coupling at first order in
$\langle Q_3^-({\bf q}_1) \rangle$, which we denote $V_c^{(1)}$
[and later the superscript 2 indicates second order in
$\langle Q_3^-({\bf q}_1) \rangle$] is
\begin{eqnarray}
V_c^{(1)} &=& \langle Q_3({\bf q}_1) \rangle \Biggl[ 
\sum_{k=1}^{n(X_2^+)]} a^{(1)}_k Z Q_2^+({\bf q}_1)_k \nonumber \\ && \ + 
\sum_{k=1}^{n(X_1^+)]} b^{(1)}_k Y Q_1^+({\bf q}_1)_k \nonumber \\
&& \  + \sum_{k=1}^{n(X_4^+)]} c^{(1)}_k Q_3^-(0) Q_4^+({\bf q}_1)_k
\Biggr] \ .
\label{VC1EQ} \end{eqnarray}
The fact that there are more than one occurrence of $Q_3({\bf q}_1)$
affects the intensity but not the number or symmetry of new modes,
$Q_2^+({\bf q}_1)_k$, $Q_1^+({\bf q}_1)_k$, and $Q_4^+({\bf q}_1)_k$.
A more complete treatment of this coupling is given in Appendix A.

We now obtain the $n[\Gamma ({\bf q}_1)]$ required in Eq. (\ref{VC1EQ}).
Using standard methods we find for wave vector ${\bf q}_1$ that the reducible
representations of the displacement vectors over each orbit have the
decomposition
\begin{eqnarray}
\Gamma [a({\bf q}_1)] &=& X_2^-({\bf q}_1) + X_3^-({\bf q}_1) + X_4^-({\bf q}_1)
\ , \nonumber \\ 
\Gamma [e({\bf q}_1)] &=& X_2^-({\bf q}_1) + X_3^-({\bf q}_1) + X_4^-({\bf q}_1)
\nonumber \\ && \ + X_1^+({\bf q}_1) + X_3^+({\bf q}_1) + X_4^+({\bf q}_1)
\nonumber \\ 
\Gamma [g({\bf q}_1)] &=& 2X_1^-({\bf q}_1) + 2X_2^-({\bf q}_1) 
+ X_3^-({\bf q}_1) \nonumber \\ && \ + X_4^-({\bf q}_1)
+ 2X_1^+({\bf q}_1) + 2X_2^+({\bf q}_1) \nonumber \\ &&  + X_3^+({\bf q}_1)
+ X_4^+({\bf q}_1) \ ,
\label{REP2EQ} \end{eqnarray}
Therefore, for example, $n[X_4^+({\bf q}_1)]$ assumes the values 0, 1, 1 for
orbits a, e, and g, respectively, leading to the results in
the second column of Table \ref{T14}.

\begin{table} [h!]
\caption{\label{T14} Number of {\it additional} optically active
modes $Q$ in the orthorhombic Cmcm phase for wave vector ${\bf q}_1$.
These are given for the Wyckoff orbits a, e, and g, using
the decompositions of Eq. (\ref{REP1EQ}) for $n[\Gamma_2^-(0)]$ and
$n[\Gamma_3^+(0)]$ and of Eq. (\ref{REP2EQ}) for $n[X_k^+({\bf q}_1)]$
and $n[X_k^-({\bf q}_1)]$.  Pol denotes polarization in the notation of
Table \ref{T9a}.  The order (in the admixture of the wave function)
is given as proportional to $\sigma^n$, where
$\sigma\equiv \langle Q_3^- ({\bf q}_1) \rangle$.  If the wave function
has an admixture of order $\sigma^n$, then the cross section has
a contribution of order $\sigma^{2n}$. TOT refers to the total for
2 a orbits, 3 e orbits and one g orbit. Note that the last row is
consistent with the last two rows of Table \ref{T9}.}
\vspace{0.2 in}
\begin{tabular} {||c || c | c | c | c || c | c | c | c | c ||}
\hline \hline
& \multicolumn{4} {c||} {Absorption \ (Odd parity) } & \multicolumn{5} {c||}
{Raman \ (Even parity)}  \\
\hline
Mode & \ $Q_4^+({\bf q}_1)$  \ & \ $Q_2^+({\bf q}_1) $ \ 
& \  $Q_1^+ ({\bf q}_1)$ \
&  \ $Q_2^-(0)$ \  &  \ $Q_1^-({\bf q}_1)$ \ 
& \ $Q_2^-({\bf q}_1)$  \
& \ $Q_3^-({\bf q}_1)$ \ & \ $Q_4^-({\bf q}_1)$ \  
&  \ $Q_3^+(0)$  \ \\
\  Pol. \ & $X$ & $Z$ & $Y$ & $X$ & $XZ$ 
& $XY$ & $X^2$, $Y^2$, $Z^2$ & $YZ$ & $Y^2-Z^2$ \\
Order & $\sigma$ & $\sigma$ & $\sigma$ & $\sigma^2$ & $\sigma$ 
& $\sigma$ & $\sigma$ & $\sigma$ &$\sigma^2$ \\
\hline \hline
RP(a) & 0 & 0 & 0 & 0 & 0 & 1 & 1 & 1 & 0  \\
RP(e) & 1 & 0 & 1 & 0 & 0 & 1 & 1 & 1 & 0  \\
RP(g) & 1 & 2 & 2 & 0 & 2 & 2 & 1 & 1 & 0  \\
\hline
TOT & 4 & 2 & 5 & 0 & 2 & 7 & 6 & 6 & 0 \\
\hline \hline
\end{tabular}
\end{table}

\begin{figure}[h!]
\begin{center}
\includegraphics[width=8.6 cm]{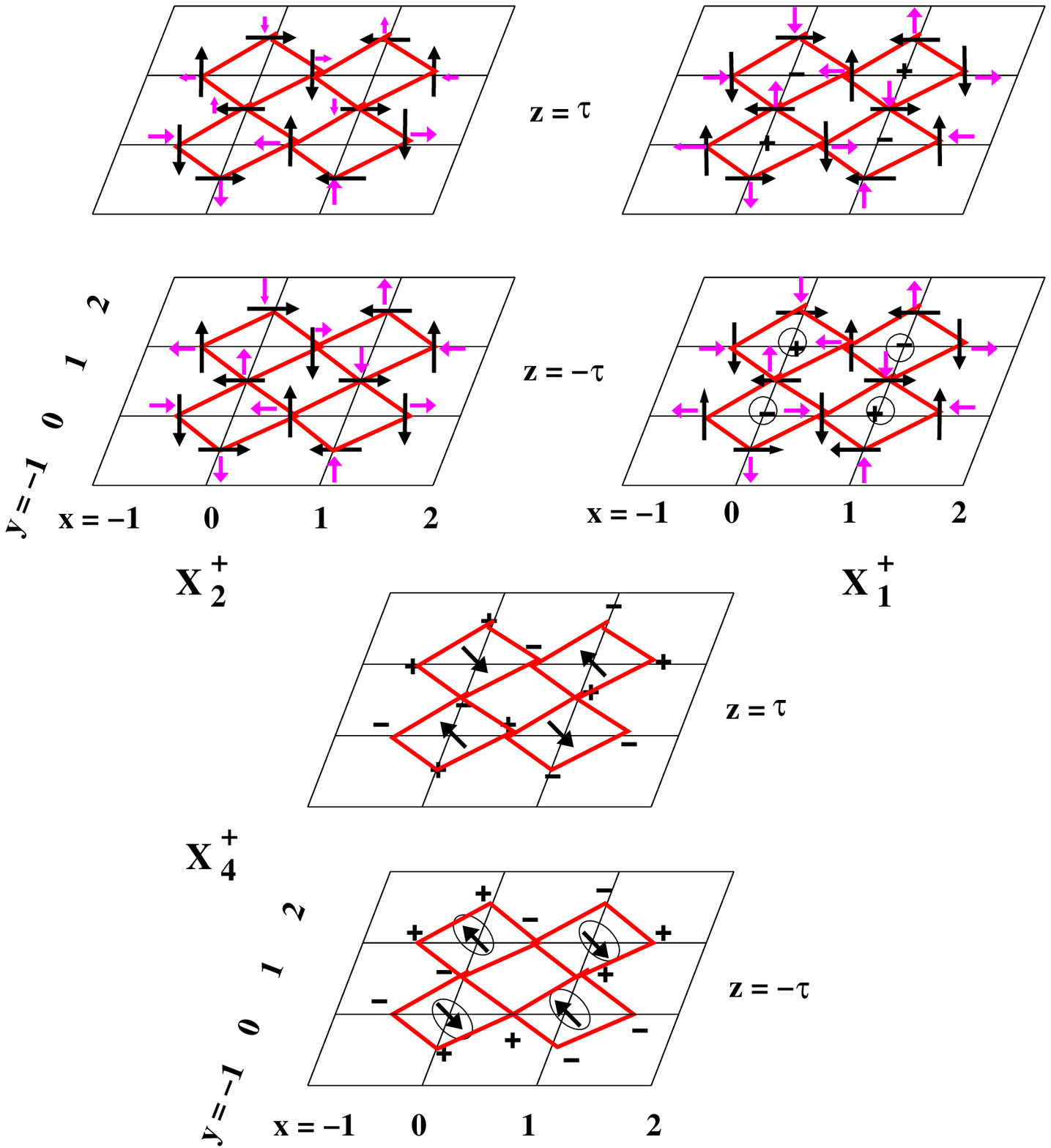}
\caption{\label{M1} (Color online)
Modes of irrep $X_2^+$, left and $X_1^+$, center,
and $X_4^+$, right, all at tetragonal wave vector ${\bf q}_1$.
These modes become absorption-active in the Cmcm phase.  The arrows represent
displacements in the $x$-$y$ plane and $+$ and $-$ are the signs of
displacements along the $z$-axis.  Here the thick-lined squares delineate the
equatorial oxygens (at $z= \pm \tau$) and the uncircled arrows and $\pm$'s
(if any) in the center of the square indicate the displacement of the top 
apical e-site oxygens in the upper layer of the bilayer at $z = \chi$ and the
circled arrows or $\pm$'s (if any) refer to the bottom apical oxygens in the
lower layer of the bilayer at $z=-\chi$.  All the e-site ions move along
the same axis as the apical oxygens, but with independent amplitudes.
There are two modes of $X_2^+$ symmetry corresponding to
independently choosing the amplitudes of the large black and small
red arrows.  If the large black arrow is dominant, this mode
is the $(0,0,\Theta)$ rotation about the tetragonal axis.  There are
five $X_1^+$ modes because the displacements of the apical oxygen,
the other two e-site ions (not shown), and the two components of the
equatorial oxygens are all independent parameters.
There are four $X_4^+$ modes because the tilting, the displacements
of the apical oxygens and of the two other e-site ions are independent
parameters.  (Some of the modes, such as $Q_1^+({\bf q}_1)$ and
$Q_4^+({\bf q}_4)$,
which grossly distort the octahedra will have relatively high frequencies.)
Without the admixtures of order $\langle Q_3^-({\bf q}_1)\rangle$
all these modes are of the wrong symmetry to be optically active.
However, the admixing, {\it e. g.} Eq. (\ref{VC1EQ}) in the Cmcm
phase makes these modes optically active.}
\end{center}
\end{figure}

We also show figures of the induced absorption modes in Fig. \ref{M1}.
The easiest way to generate a mode of a given symmetry ($X_k^+$ or $X_k^-$)
is to consider each Wyckoff orbit in turn.  For a given orbit assign one site
an arbitrary displacement vector $(u_x, u_y, u_z)$, then generate the
displacements of the other sites using the characters of Table
\ref{IRREP}. In doing this one will find that one or more of the initial
displacements can not be nonzero when one includes the restriction that
the wave function have the given wave vector, in this case 
${\bf q}_1$.

Now we turn to the case $n=2$, for which
\begin{eqnarray}
{\cal I} U_2 (0) &=& - U_2(0) \ , \ \
m_z U_2(0) = U_2(0) \ , \nonumber \\ 
m_d U_2(0) &=& U_2(0) \ , \ \
{\cal I} V_2(0)  = - V_2(0) \ , \nonumber \\
m_z V_2(0) &=& V_2(0) \ , \ \ \ \
m_d V_2(0) =  - V_2(0) \ ,  \nonumber \\
{\cal I} W_2 (0) &=& - W_2(0) \ , \ \ \ \
m_z W_2(0) = - W_2(0) \ , \nonumber \\
m_d W_2(0) &=& W_2(0) \ . \nonumber
\end{eqnarray}
The solutions for $U_2(0)$ and $V_2(0)$ do not produce new modes but
instead add to the intensity of already existing modes.  New modes arise
from $W_2(0)=Q_2^-(0)$, so that the relevant second order coupling
perturbation is
\begin{eqnarray}
V_c^{(2)} &=& \langle Q_3^-({\bf q}_1) \rangle^2 \sum_{k=1}^{n[\Gamma_2^-(0)]}
c^{(2)}_k Q_3^-(0) Q_2^-(0)_k \ , \nonumber
\end{eqnarray}
where ${n[\Gamma_2^-(0)]}$ is given in Eq. (\ref{REP1EQ}).
We summarize our results in Table \ref{T14}.
Note that all the newly induced modes have absorption intensity
proportional to $\langle Q_3^-({\bf q}_1)\rangle^2$ only because
the low symmetry sites of I4/mmm (which allow the OP $Q_2^-(0)$
to appear) are not occupied for CMO or CTO: $n[\Gamma_2^-(0)]=0$ for
CMO or CTO. But an argument equivalent to what we have given here
would be needed to establish this result.

%

\subsubsection{\it Induced Raman Modes}

Here we consider the admixing of the Raman active modes (see Table \ref{T4})
from irreps $\Gamma_n^+$, with $n=1,2,4,5$:
\begin{eqnarray}
V_C &=& \sum_n \langle Q_3^-({\bf q}_1 \rangle^n \Bigg[ 
a_n Z^+ T_n + b_n Q_4^+(0) U_n \nonumber \\ && \ + c_n Q_2^+(0) V_n
+ d_n Q_1^+(0) W_n + e_n Y^+ S_n \Biggr] \ , \nonumber
\end{eqnarray}
where $T_n$, $U_n$, $V_n$, $W_n$, and $S_n$ are the modes which are
Raman active by virtue of their coupling proportional to
$\langle Q_3^- \rangle({\bf q}_1)^n$ to modes of the tetragonal
system which are Raman active.
For $n=1$ invariance with respect to ${\cal I}$, $m_d$, and $m_z$ leads to
\begin{eqnarray}
{\cal I} S_1({\bf q}_1) &=& - S_1({\bf q}_1) \ , \ \ \ \ 
m_d S_1({\bf q}_1) = S_1({\bf q}_1) \ , \nonumber \\
m_z S_1({\bf q}_1) &=& - S_1({\bf q}_1) \ , \ \ \ \
{\cal I} T_1({\bf q}_1) = - T_1({\bf q}_1) \ , \nonumber \\
m_d T_1({\bf q}_1) &=& - T_1({\bf q}_1) \ , \ \ \ \
m_z T_1({\bf q}_1) = - T_1({\bf q}_1) \nonumber \\ 
{\cal I} U_1({\bf q}_1) &=& - U_1({\bf q}_1) \ , \ \ \ \ 
m_d U_1({\bf q}_1) = - U_1({\bf q}_1) \ , \nonumber \\
m_z U_1({\bf q}_1) &=& U_1({\bf q}_1) \ , \ \ \ \
{\cal I} V_1({\bf q}_1) = - V_1({\bf q}_1) \ , \nonumber \\
m_d V_1({\bf q}_1) &=& V_1({\bf q}_1) \ , \ \ \ \
m_z V_1({\bf q}_1) = V_1({\bf q}_1) \ , \nonumber \\
{\cal I} W_1({\bf q}_1) &=& - W_1({\bf q}_1) \ , \ \ \ \ 
m_d W_1({\bf q}_1) = - W_1({\bf q}_1) \ , \nonumber \\
m_z W_1({\bf q}_1) &=& W_1({\bf q}_1) \ . \nonumber
\end{eqnarray}
So
\begin{eqnarray}
S_1({\bf q}_1) &=& Q_2^- ({\bf q}_1) \ , \
T_1({\bf q}_1) = Q_1^-({\bf q}_1) \ ,  \ U_1({\bf q}_1)
= Q_3^-({\bf q}_1)\ , \nonumber \\
V_1({\bf q}_1)&=&Q_4^-({\bf q}_1)\ ,
\ \ \ \ W_1({\bf q}_1)=Q_3^-({\bf q}_1) \ . \nonumber
\end{eqnarray}
Thus 
\begin{eqnarray}
V_C^{(n=1)} &=& \langle Q_3^-({\bf q}_1) \rangle \Biggl[ 
\sum_{k=1}^{n(X^-)} a^{(1)}_k Z^+ Q_1^-({\bf q}_1)_k
\nonumber \\ && \
+ \sum_{k=1}^{n(X_3^-)} b^{(1)}_k Q_4^+(0) Q_3^-({\bf q}_1)_k
\nonumber \\ && \
+ \sum_{k=1}^{n(X_4^-)} c^{(1)}_k Q_2^+(0) Q_4^-({\bf q}_1)_k
\nonumber \\ && \
+ \sum_{k=1}^{n(X_3^-)} d^{(1)}_k Q_1^+(0) Q_3^-({\bf q}_1)_k
\nonumber \\ && \
+ \sum_{k=1}^{nX_2^-)} e^{(1)}_k Y^+ Q_2^-({\bf q}_1)_k
\Biggr] \ , \nonumber
\end{eqnarray}
where the sums over the number of appearances of the Raman active tetragonal
modes is omitted, in analogy with Eq. (\ref{VC1EQ}).
The fact that $Q_4^+(0)$ and $Q_1^+(0)$ are both coupled to 
$X_3^-({\bf q}_1)_k$ means that each $X_{3,k}$ mode has two independent
admixtures, one, $Q_4^+(0)$, giving a contribution to the Raman tensor
$a(\hat i \hat j + \hat j \hat i)= a(\hat K \hat K - \hat J \hat J)$
and the other, $Q_1^+(0)$, giving a contribution to the Raman tensor
$b\hat k \hat k + c(\hat i \hat i + \hat j \hat j)
= b \hat I \hat I + c(\hat J \hat J + \hat K \hat K)$.
As a result, the $X_{3,k}^-({\bf q}_1)$ modes have a Raman tensor 
$b\hat I \hat I + (c-a)(\hat J \hat J)
+ (c+a)\hat K \hat K$, so that $X_{3,k}^-({\bf q}_1)$
has a Raman polarization we denote $(X^2,Y^2,Z^2)$.

For $n=2$ invariance with respect to ${\cal I}$, $m_d$, and $m_z$ leads to
\begin{eqnarray}
{\cal I} S_2(0) &=& S_2(0) \ , \ \ m_d S_2(0) = -S_2(0) \ , \nonumber \\
m_z S_2(0) &=& -S_2(0) \  , \ \ {\cal I} T_2(0) = T_2(0) \ , \nonumber \\
m_d T_2(0) &=& T_2(0) \ , \ \ m_z T_2(0) = - T_2(0) \nonumber \\ 
{\cal I} U_2(0) &=& U_2(0) \ , \ \ m_d U_2(0) = U_2(0) \ , \nonumber \\
m_z U_2(0) &=& U_2(0) \ , \ \ {\cal I} V_2(0) = V_2(0) \ , \nonumber \\
m_d V_2(0) &=& -V_2(0) \ , \ \ m_z V_2(0) = V_2(0) \ , \nonumber \\
{\cal I} W_2(0) &=& W_2(0) \ , \ \ \ \ m_d W_2(0) = W_2(0) \ , \nonumber \\
m_z W_2(0) &=& W_2(0) \ . \nonumber
\end{eqnarray}
So
\begin{eqnarray}
S_2 &=&Y_O^+(0) \ , \ \ T_2= Z^+\ , \nonumber \\
U_2 &=& \ \alpha Q_1^+(0) + \beta Q_4^+ (0) \ , \nonumber \\
V_2 &=& \ \alpha' Q_2^+(0) +\beta' Q_3^+ (0) \ , \nonumber \\
W_2 &=& \ \alpha^{\prime \prime} Q_1^+ (0) + \beta^{\prime \prime} Q_4^+ (0)
\ .  \nonumber \end{eqnarray}
The only new irrep here is $\Gamma_3^+ (0)$ with OP $Q_3^+(0)$.
So we have the results of
Table \ref{T14}. As was the case for the absorption cross section,
all the newly induced modes have their Raman scattering cross section
proportional to $\langle Q_3^-({\bf q}_1)\rangle^2$ only because
the low symmetry sites of I4/mmm (which allow the OP $Q_3^+(0)$
to appear) are not occupied for CMO or CTO. We emphasize that this result is
not a general result for the space group I4/mmm distorting into space
group Cmcm.

\begin{figure}[h!]
\begin{center}
\includegraphics[width=8.6 cm]{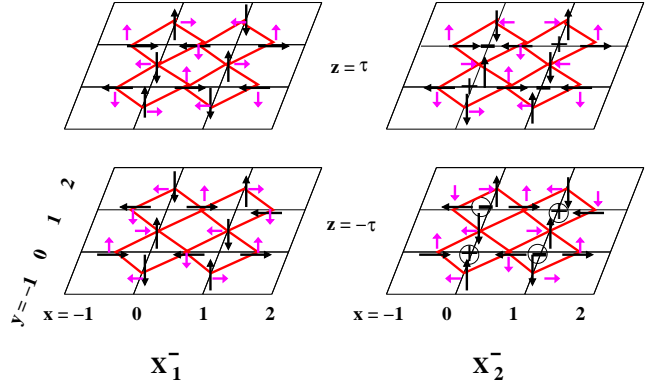}
\caption{\label{M2} (Color online)
As Fig. \ref{M1} except for modes of irrep $X_1^-$,
left and $X_2^-$, right, both at tetragonal wave vector ${\bf q}_1$.
These are the modes which become Raman active (see Table \ref{T14})
due to admixtures of tetragonal Raman active zone center modes.
In mode $X_1^-$ the sites on the $z$ axis do not move, so
there are two modes of this symmetry corresponding to the choice of
the amplitudes of the $x$ and $y$ displacements. There are 7 modes
of symmetry $X_2^-({\bf q}_1)$ because there are two ways to choose
the amplitude of the arrows (the displacements of the equatorial oxygens)
and one way to independently choose the amplitudes of the displacements
along the $z$-axis for each of the 2 a orbits and for each of the 3 e orbits
of such sites.
These breathing modes grossly distort the octahedra and therefore have
relatively high frequency.}
\end{center}
\end{figure}

\begin{figure}[h!]
\begin{center}
\includegraphics[width=8.6 cm]{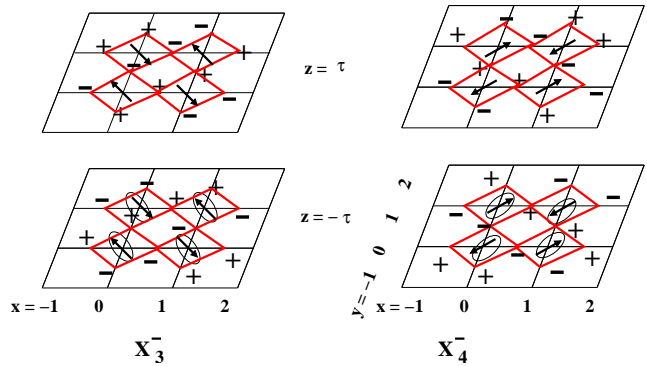}
\caption{\label{M3} (Color online)
As Fig. \ref{M2} except for modes of irrep $X_3^-$,
left and $X_4^-$, right, both at tetragonal wave vector ${\bf q}_1$.
The equatorial oxygens (g orbits) tilt around the [110] direction for
irrep $X_3^-({\bf q}_1)$
and around the [1$\overline 1$0] direction for $X_4^-({\bf q}_1)$, in
each case with a single amplitude. The other ions on the $z$-axis move
along the [1$\overline 1$0] axis for $X_3^-({\bf q}_1)$ and along the
[110] axis for $X_4^-({\bf q}_1)$, in each case with an independent
amplitude. Thus for both irreps we have 6 modes (one for each
orbit). In the notation of Refs. \onlinecite{ROTA} and \onlinecite{ROTB}
$Q_3^-({\bf q}_1)$
is a $(\Phi,\Phi,0)$ rotation and $Q_4^-({\bf q}_1)$ is a $(\Phi,-\Phi,0)$
rotation. The wave vector regulates how one bilayer is structured relative
to the adjacent bilayer.  So these two modes have slightly different
frequencies because although their intrabilayer interaction energies are
the same, their interbilayer interaction energies are different.}
\end{center}
\end{figure}

In Figs. \ref{M2} and \ref{M3} we show the
new Raman modes in the Cmcm phase.

\vspace{0.2 in}

\vspace{0.2 in}

\subsubsection{Temperature Dependence of Mode Energies
and Absorption Intensities}

\vspace{0.2 in}

Near the phase transition at temperature $T=T_>$ where the order parameter
$\langle Q_3^-({\bf q}_1) \rangle$ becomes nonzero one has
\begin{eqnarray}
\omega_n(T) &=& \omega_n^{(0)} (T) + \delta \omega_n (T) \ , \nonumber \\
I_n(T) &=& I_n^{(0)} (T) + \delta I_n (T) \ , \ \ \ \
\label{T3EQ} \end{eqnarray}
where the first term is the smooth temperature dependence and the
second term is the anomalous term at the phase transition which
is only nonzero for $T<T_>$ and will involve powers of
$\langle Q_3^-({\bf q}_1) \rangle \equiv \sigma \sim |T_>-T|^\beta$,
where the order parameter exponent $\beta$ is 1/2 in mean field theory,
but fluctuations will cause it to be less than 1/2.

\begin{figure}[h!]
\begin{center}
\includegraphics[width=8.6 cm]{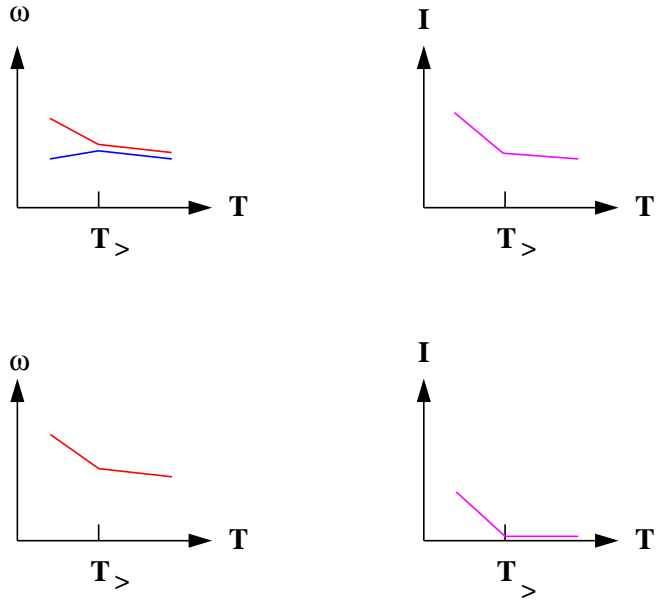}
\caption{\label{MODES} (Color online)
RP modes.  Left panels: mode energy versus temperature,
where $T_>$ is the temperature at which $\langle Q_3^-({\bf q}_1)\rangle$
becomes nonzero.  Right panels: the mode intensity versus temperature.
Upper left: removal of degeneracy of the $\Gamma_5^\pm$ modes. Lower left:
temperature dependence of a nondegenerate mode energy.  Upper right:
temperature dependence of the intensity of a nondegenerate mode which
is absorption allowed in the tetragonal phase.  Lower right:
temperature dependence of nondegenerate mode intensity for modes which
have zero intensity in the tetragonal phase.  In all cases, the discontinuity
in slope is drawn assuming the mean-field result, $\beta=1/2$, for
the order parameter exponent.  Otherwise, the lower temperature
additional contribution is of order $|T_>-T|^{2 \beta}$.}
\end{center}
\end{figure}

First consider the mode energies.  For all the {\it nondegenerate}
${\bf q}=0$ modes, the perturbation of the form given in Eq. (\ref{VVEQ}),
will lead to an energy shift from terms with $n=1$, which, in second order
perturbation theory, gives a contribution to $\delta \omega (T)$ of order
$\sigma^2 \sim|T_>-T|^{2 \beta}$.
 For the {\it degenerate} $\Gamma_5^\pm(0)$ modes,
$\delta \omega (T)$ has a contribution from first order perturbation
theory, but in that case, since $b_1$ of Eq. (\ref{EQHAM}) vanishes,
the contribution to $\delta \omega (T)$ is again of order 
$\sigma^2 \sim |T_>-T|^{2 \beta}$.
These results are shown in the left-hand panels of Fig. \ref{MODES}.
It should also be noted that the mode energies can be modified by
quartic terms in the Hamiltonian which we have not so far considered
and which are of the form[\onlinecite{RABE,ABHPRB}]
\begin{eqnarray}
V_4 &=& \sum_n d_n U_n^2 Q_3^-({\bf q}_1)^2 \nonumber
\end{eqnarray}
which will also lead to a contribution to $\delta \omega (T)$ of order
$\sigma^2 \sim |T_>-T|^{2 \beta}$. It is plausible, but not a certainty,
that the sign of the anomalous contribution would cause a stiffening of the
modes as the temperature is reduced as is depicted in lower left-hand panel
of Fig. \ref{MODES}. If $V_4$ were to be neglected, then one would have a
direct correlation between 
$\delta I_n(T)$ and $\delta \omega_n(T)$.[\onlinecite{FN4}]
 
Now consider the mode intensities in the presence of admixtures
written in Eq. (\ref{MIX}).  The standard scenario in
perturbation theory is that a perturbation in first order introduces
new components into the wave function without changing the unperturbed
component of the wave function.  In second order perturbation theory
the amplitude of the unperturbed component of the wave function can
be modified. Here, this tells us that if the mode $U_n$ is already
dipole allowed, then the intensity will be modified at order
$\sigma^2$.  If the mode $U_n$ is not allowed in the absence of the
perturbation then its intensity will be proportional to the
square of the amplitude of the admixture.  Furthermore, new modes
that appear at first order in $\sigma$ will have intensity of
the form of Eq. (\ref{T3EQ}) with $I_n^{(0)}(T)=0$ and
$\delta I (T) \sim \sigma^2 \sim |T_>-T|^{2 \beta}$.  New modes
that only appear at second order in $\sigma$ will have intensity of
the form of Eq. (\ref{T3EQ}) with $I_n^{(0)}(T)=0$ and
$\delta I (T) \sim \sigma^4 \sim |T_>-T|^{4 \beta}$.
From Table \ref{T14} we see that for CMO or CTO there
are no such modes that only appear at order $\sigma^2$.  This
is a special property of the RP system.  Had we had a more general
tetragonal system with low symmetry sites, then there would be
modes which only appear at order $\sigma^2$ and which would have
$\delta I (T) \sim \sigma^4 \sim |T_>-T|^{4 \beta}$.  Our
results are summarized in the right-hand panels of Fig. \ref{MODES}.

\section{Cmc2$_1$ Phase}

\subsection{Analysis based on Cmc2$_1$ symmetry}

Since Cmc2$_1$ does not have a center of inversion symmetry, parity is not a
good quantum number and Raman and absorption modes are mixed.  In Cmc2$_1$ 
the general orbit is half as large as for Cmcm.
However, the unit cell then contains twice as many orbits. The
sites in Cmc2$_1$ are given in Table \ref{T15} and the character
table for ${\bf q}=0$ is given in Table \ref{T16}.

\begin{table} [h!]
\caption{\label{T15} Operators for Cmc2$_1$.}
\vspace{0.2 in}
\begin{tabular} {||c c ||}
\hline \hline
$E=(x,y,z)$ &\ \ $2_z=(\overline x , \overline y, 1/2+z)$\ \ \\
\ \ $m_x=(\overline x,y,z)$\ \ & $m_y=(x, \overline y, 1/2+z)$ \ \ \\
\hline \hline
\end{tabular}
\end{table}

\begin{table}
\caption{\label{T16} Character table for the little group at 
${\bf q}=0$ for the Cmc2$_1$ lattice. The last three columns are the
characters for the various tetragonal orbits on the Cmc2$_1$ structure.
Orthorhombic coordinates are used throughout this table.
${\bf P}$ gives the nonzero component of the dipole moment operators
for the irrep in question.  The Raman tensors ${\bf R}$ are
${\bf R}(\Gamma_1)=d \hat I \hat I + e \hat J \hat J + f \hat K \hat K$,
${\bf R}(\Gamma_2)=g(\hat J \hat K + \hat K \hat J)$,
${\bf R}(\Gamma_3)=h(\hat I \hat J + \hat J \hat I)$, and
${\bf R}(\Gamma_4)=i(\hat I \hat K + \hat K \hat I)$.}
\vspace{0.2 in}
\begin{tabular} {|| c || c | c | c | c ||c | c | c ||}
\hline \hline
Operator &\ \ $\Gamma_1$\ \ & \ \ $\Gamma_2$\ \ &\ \ $\Gamma_3$\ \ & \ \
$\Gamma_4$\ \ &\ \ a \ \ & \ \ e \ \ & \ \ g \ \ \\
\hline
$(X,Y,Z)$ & $1$ & $1$ & $1$ & $1$ & 6 & 12 & 24 \\
\ \ $(\overline X , Y,Z)$ 
& $1$ & $1$ & $-1$ & $-1$ & 2 & 0 & 0 \\
\ \ $(\overline X , \overline Y ,1/2+Z)$ \ \
& $1$ & $-1$ & $1$ & $-1$ & 0 & 0 & 0 \\
$(X , \overline Y , 1/2+Z)$
& $1$ & $-1$ & $-1$ & $1$ & 0 & 0 & 0 \\
\hline
${\bf P}$ & $a\hat K$ & $b\hat J$ & None & $c\hat I$ & & & \\
\hline
\hline \hline
\end{tabular}
\end{table}

Using the data in Table \ref{T16} we find that
\begin{eqnarray}
\Gamma(a_{Cmc2_1}) &=& 2 \Gamma_1 + 2\Gamma_2 + \Gamma_3 + \Gamma_4\ ,
\nonumber \\
\Gamma(e_{Cmc2_1}) &=& 3 \sum_{n=1}^4  \Gamma_n \ , \
\Gamma(g_{Cmc2_1}) = 6 \sum_{n=1}^4 \Gamma_n \ ,
\label{CMC2EQ} \end{eqnarray}
and the dipole moments and Raman tensors of each irrep are given in Table
\ref{T16}.
Note that once we have identified the absorption modes, the identification of
the Raman active modes follows, except for those of irrep $\Gamma_3$.
Using Eq. (\ref{CMC2EQ}) we find the number of optically active modes in
the Cmc2$_1$ phase given in Table \ref{T19}.

\begin{table}
\caption{\label{T19} As Table \ref{T16}.
Optically active modes of the Cmc2$_1$ phase.
TOT gives the total number of modes of each symmetry found using
Eq. (\ref{CMC2EQ}) and Table \ref{T16}.  We copy the number of
modes in the Cmcm phase from Table \ref{T9} and Addl gives the
number of new modes which appear when the Cmc2$_1$ phase is
reached, as given in Eqs. (\ref{UVWEQ}) and (\ref{R1R2EQ}).
Here R refers to Raman active modes and P to dipole-active modes.
This entry includes only the additional modes we explicitly
calculated for the dipole active modes and for the Raman active modes
which are {\it not} dipole active. Since the total number of modes
is given for both Cmcm and Cmc2$_1$, one can deduce that, for instance,
the number of additional Raman active modes with Raman tensor
$i(\hat I \hat K + \hat K \hat I)$ is 17-7=10.}

\vspace {0.2 in}
\begin{tabular} {|| c || c | c | c | c ||}
\hline \hline
Irrep&\ \ $\Gamma_1$\ \ & \ \ $\Gamma_2$\ \ &\ \ $\Gamma_3$\ \ & \ \
$\Gamma_4$ \ \ \\
\hline
${\bf P}$ & $a \hat K$ & $b \hat J$ & None & $c \hat I$ \\
\hline
TOT & 19 & 19 & 17 & 17 \\
\hline
Cmcm & 10(R), 9(P) & 7(R), 12(P) & 12(R) & 7(R), 10(P) \\
Addl & 10(P) & 7(P) & 5(R) & 7(P) \\
\hline \hline
\end{tabular}
\end{table}

\subsection{Analysis based on lowering of Cmcm symmetry}

Now we perform the analysis of mode symmetry using the mixing of modes
in terms of the OP's present in the Cmc2$_1$ phase, namely $Q_3^-({\bf q}_1)$,
$Z\equiv [\Gamma_{5,1}^- + \Gamma_{5,2}^-]/\sqrt 2$, and $Q_2^+({\bf q}_1)$,
where we assume the OP's of the family of ${\bf q}_1$. (Afterwards
the analogous
terms associated with ${\bf q}_2$ can be obtained by applying the four-fold
rotation ${\cal R}_4$ to the coupling we construct for ${\bf q}_1$.)
At the lower transition we can mix using all the condensed order parameters,
$Q_3^-({\bf q}_1)$, $Q_2^+ ({\bf q}_1)$, and $Z$.
The first question is: can we induce
$Y \equiv Q_{5,1}^-(0) - Q_{5,2}^-(0)$?
The answer is no.   Note that $\sigmav_d Z =Z$ and $\sigmav_d Y=-Y$.
Also $\sigmav_d Q_3^-({\bf q}_1)= -\sigmav_d Q_3^-({\bf q}_1)$ and
$\sigmav_d Q_2^+({\bf q}_1) =-  Q_2^+({\bf q}_1)$.  But for wave
vector conservation we must have a total even number of powers of these
two operators at ${\bf q}_1$.  Hence there can be no term in the free energy
which is linear in $Y$.  So $Y$ is not induced by the other order parameters. 

\subsubsection{Induced Absorption Modes}

So optically active absorption modes are admixed by the perturbation
\begin{eqnarray}
&& V({\bf q}_1) = \sum_{k,l,m} \langle Q_3^-({\bf q}_1) \rangle^k
\langle Q_2^+({\bf q}_1) \rangle^l
\langle Z \rangle^m
\nonumber \\ &\times& \Biggl( a_{klm} U_{klm} Z
+ b_{klm} V_{klm} Y + c_{klm} W_{klm} Q_3^-(0) \Biggr) \ ,
\nonumber \end{eqnarray}
where again the $U$'s, $V$'s, and $W$'s are the single phonon operators
which will become optically active due to this mechanism and which we wish
to identify.
We insert the angle brackets to emphasize that these higher order terms give 
rise to an effective bilinear coupling of the type that mixes modes.  
In principle any perturbation has to be invariant under all the
symmetries of the tetragonal phase.  However, applying a four-fold
rotation takes ${\bf q}_1$ into ${\bf q}_2$ (whose order parameters are zero)
and $\langle Z \rangle$ into $\langle Y \rangle$  (which is also zero).
Thus we require that $V({\bf q}_1)$ be invariant under only $m_d$,
$m_z$, and ${\cal I}$, in which case
\begin{eqnarray}
m_dU_{klm}&=&(-)^{k+l+1}U_{klm} \ , \ \
m_dV_{klm}=(-)^{k+l}V_{klm} \ , \nonumber \\
m_dW_{klm}&=&(-)^{k+l+1}W_{klm} \ , \ \
m_zU_{klm}=U_{klm} \ , \nonumber \\
m_zV_{klm}&=&V_{klm} \ , \ \
m_zW_{klm}=-W_{klm} \nonumber \\
{\cal I}U_{klm}&=&(-)^{k+m+1}U_{klm} \ , \ \
{\cal I}V_{klm}=(-)^{k+m+1}V_{klm} \ , \nonumber \\
{\cal I}W_{klm}&=&(-)^{k+m+1}W_{klm} \ . \nonumber
\end{eqnarray}

\begin{table} [h!]
\caption{\label{T13} Independent sets of indices for the operators
$U$, $V$, or $W$ (which depend on $(-)^{k+m}$ and $(-)^{k+l}$).
``Cmcm" indicates equality to cases
previously considered for Cmcm.  The equal sign indicates equality in view
of $\langle Z \rangle \propto \langle Q_3^-({\bf q}_1)\rangle
\langle Q_2^+ ({\bf q}_1) \rangle$.  Cases which lead to identical 
operators with different prefactors are indicated by "$\propto$." 
Cases which produce ``new" operators are indicated by $*$.  }
\vspace{0.2 in}
\begin{tabular} {||c || c c c c c c c ||}
\hline \hline
$k=$ & 0 & 0 & 1 & 1 & 1 & 0 & 1 \\
$l=$ & 0 & 1 & 0 & 1 & 0 & 1 & 1 \\
$m=$ & 1 & 0 & 0 & 0 & 1 & 1 & 1 \\
\hline
\ \ Case \ \ & \ \ A \ \ & \ \ B \ \ & \ \ C \ \
& \ \ D \ \ & \ \ E \ \ & \ \ F \ \ & \ \ G \\
\hline
New? & $*$ & $*$ & Cmcm & $=$A & $\propto$ B & $=$C & $\propto$ C \\ 
\hline \hline
\end{tabular}
\end{table}

\noindent
We use the result of Ref. \onlinecite{ABHCF} that, due to the
cubic coupling between the order parameters, one has
\begin{eqnarray}
\langle Z \rangle = a \langle Q_2^+({\bf q}_1) \rangle
\langle Q_3^-({\bf q}_1) \rangle \ ,
\end{eqnarray}
where $a$ is a constant.
Note the following: a) increasing $k$, $l$, or $m$ by two does not induce
an additional operator and b) solutions for $U$, $V$, and $W$, depend on
$k+l$ and $k+m$. Therefore we have the results of Table \ref{T13} of
which we only consider cases A, B, and E.  We find that
\begin{eqnarray}
U_{001} &=& c Q_1^+(0) + d Q_4^+(0) \ , \ \
V_{001} = c' Q_2^+(0) + d' Q_3^+(0) \ , \nonumber \\
W_{001} &=& Z^+ \ , \ \
U_{101} = U_{010} = Q_3^-({\bf q}_1 )  \ , \nonumber \\
V_{101} &=& V_{010} = Q_4^-({\bf q}_1 )  \ , \ \
W_{101} = W_{010} = Q_1^-({\bf q}_1 )  \ .
\label{UVWEQ} \end{eqnarray}
From these three cases we get (for wave vector ${\bf q}_1$)
\begin{eqnarray}
&& V({\bf q}_1) = \langle Q_2^+ ({\bf q}_1) \rangle
\langle Q_3^- ({\bf q}_1) \rangle \Biggl\{ a_1 Q_3^-(0) Z^+
+ Z [ \nonumber \\ && \ a_2 Q_1^+ (0) + a_3 Q_4^+(0)]
+ Y [ a_4 Q_2^+ (0) + a_5 Q_3^+(0)] \Biggr\} \nonumber \\ && \
+ \langle Q_2^+ ({\bf q}_1) \rangle \Biggl\{ a_6 Z Q_3^-({\bf q}_1) +
a_7 Y Q_4^- ({\bf q}_1) \nonumber \\ && \ 
+ a_8 Q_3^-(0) Q_1^-({\bf q}_1) \Biggr\}
+ \langle Q_2^+({\bf q}_1) \rangle
\langle Q_3^-({\bf q}_1) \rangle^2 \Biggl\{ \nonumber \\ &&
a_9 Z Q_3^-({\bf q}_1) + a _{10} Y Q_4^- ({\bf q}_1)
+ a_{11} Q_3^-(0)Q_1^-({\bf q}_1) \Biggr\} \ . \nonumber
\\ \label{EQ46} \end{eqnarray}

By applying ${\cal R}_4$ to this we get
\begin{eqnarray}
&& V({\bf q}_2) = \langle Q_2^+ ({\bf q}_2) \rangle
\langle Q_3^- ({\bf q}_2) \rangle \Biggl\{ a_1 Q_3^-(0) Y^+
+ Y [ \nonumber \\ && \ a_2 Q_1^+ (0) - a_3 Q_4^+(0)]
+ Z [ a_4 Q_2^+ (0) - a_5 Q_3^+(0)] \Biggr\} \nonumber \\ && \
+ \langle Q_2^+ ({\bf q}_2) \rangle \Biggl\{ a_6 Y Q_3^-({\bf q}_2) +
a_7 Z Q_4^- ({\bf q}_2) \nonumber \\ && \ 
+ a_8 Q_3^-(0) Q_1^-({\bf q}_2) \Biggr\}
+ \langle Q_2^+({\bf q}_2) \rangle
\langle Q_3^-({\bf q}_2) \rangle^2 \Biggl\{ \nonumber \\ &&
a_9 Y Q_3^-({\bf q}_2) + a _{10} Z Q_4^- ({\bf q}_2)
+ a_{11} Q_3^-(0)Q_1^-({\bf q}_2) \Biggr\} \ . \nonumber
\\ \label{EQ47} \end{eqnarray}

\begin{table} [h!]
\caption{\label{T17} {\it Additional} optically active modes (which we 
refer to as `new') which appear in the Cmc2$_1$ phase for wave vector
${\bf q}_1$ from Eqs. (\ref{UVWEQ}) and (\ref{R1R2EQ}). 
The values of $n[(\Gamma(0)]$ for each orbit are given in Eq. (\ref{TETIRREP})
and those of $n[(\Gamma({\bf q})]$ are given in Eq. (\ref{REP2EQ}).
TOT is the total number of modes from 2a orbits, 3e orbits, and one g orbit.
The values of ${\bf P}$ (and the associated values of the Raman tensor
${\bf R}$) are from $Z$, $Y$ and $Q_3^-(0)$ in Eq.
(\ref{EQ46}). The values of the Raman tensor ${\bf R}$ for $Q_4^-(0)$ is
inherited from $Y^+$ in Eq. (\ref{EQ48}).  In the last row the entry
1 indicates that the Raman tensor is ${\cal R} = a \hat I \hat I
+ b \hat J \hat J + c \hat K \hat K$.}
\vspace{0.2 in}
\begin{tabular} {||c | c c c c c c c | c c c c || }
\hline \hline
& \multicolumn{7} {c|} {New zero wave vector modes} & 
\multicolumn{4} {c|} {New modes for wave vector ${\bf q}_1$} \\
\hline
Orbit & $Q_1^+$(0) & $Q_4^+(0)$ & $Q_2^+(0)$ & $Q_3^+(0)$
& $Z^+$ & $Q_1^-(0)$ & $Q_4^-(0)$ & $Q_3^-({\bf q}_1)$ & $Q_4^-({\bf q}_1)$ &
$Q_1^-({\bf q}_1)$ & $Q_3^+({\bf q}_1)$ \\
\hline
a  & 0 & 0  & 0 & 0 & 0 & 0 & 0 & 1 & 1 & 0 & 0  \\
e  & 1 &  0 & 0 & 0 & 1 & 0 & 0 & 1 & 1 & 0 & 1  \\
g  & 1 &  0 & 1 & 0 & 2 & 0 & 1 & 1 & 1 & 2 & 1  \\
\hline
TOT & 4 & 0 & 1 & 0 & 5 & 0 & 1 & 6 & 6 & 2 & 4  \\
\hline
{\bf P} & Z & Z & Y & Y & X & none & none & Z& Y & X & none \\
{\bf R} & 1 & 1 & YZ& YZ& XZ& none & XY   & 1& YZ& XZ& XY \\
\hline \hline
\end{tabular}
\end{table}

\subsubsection{Induced Raman Modes}

Finally, we carry out the same analysis to determine the new Raman modes of
polarization $X_OY_O$, since these modes are not dipole active and hence
are not accessible to the previous calculation. To get
polarization $X_OY_O=z_t(x_t-y_t)$ we must induce the mode 
$Y^+$ of irrep $\Gamma_5^+$. We
use the results of Table \ref{T13} to restrict our attention to the
coupling for the cases $k=l=m-1=0$ (together with $k=l=m+1=1$) and
$k=l-1=m=0$ (together with $k=l+1=m=1$) so that
\begin{eqnarray}
&& V = R_1 Y^+ [\alpha \langle  Z \rangle + \beta 
\langle Q_2^+({\bf q}_1) \rangle \langle Q_3^-({\bf q}_1) \rangle ]
\nonumber \\
&& + R_2 Y^+ [\gamma \langle  Q_2^+({\bf q}_1) \rangle + \delta 
\langle Q_3^-({\bf q}_1) \rangle \langle Z \rangle ] \ ,
\label{EQ48} \end{eqnarray}
where $R_k$ are the single phonon operators which we determine by
requiring that $V$ be an invariant under ${\cal I}$, $m_d$, and $m_z$.
we find that
\begin{eqnarray}
R_1 &=& Q_1(0) \ {\rm or} \ Q_4^- (0) \ , \hspace{0.2 in}
R_2 = Q_3^+({\bf q}_1) \ .
\label{R1R2EQ} \end{eqnarray}
The number of modes we admix into $R_1$ is equal to the number of times
either $\Gamma_1^-(0)$ or $\Gamma_4^-(0)$ appears, which from Eq.
(\ref{TETIRREP}) is one for each g orbit or once in all.  The
number of modes we admix into $R_2$ is equal to the number of times
irrep $\Gamma_3^+({\bf q}_1)$ appears, which from Eq. (\ref{REP2EQ})
is one for each e orbit and one for each g orbit, or four in all.
So there are five additional $X_OY_O$ Raman modes induced in Cmc2$_1$.

We could give figures showing the displacements in the new optically
active modes which appear in the Cmc2$_1$ phase.  Actually, the new
modes at wave vector ${\bf q}_1$ have already been shown in Figs.
\ref{M1}, \ref{M2}, and \ref{M3}, but now many of these modes are both
absorption active and Raman active.  We leave it to the reader to
draw the new modes at zero wave vector using the same technique as
used above.

\subsubsection{\it Summary of Optically Active Phonons.}

In Table \ref{ACTIVE} we list the irreps of the tetragonal phase
which are active or become active due to the condensation of order
parameters $Q_3^-({\bf q})$, $Q_2^+({\bf q})$, and $\Gamma_5^-(0)$.
However, one can show that the intensities for wave vector ${\bf q}_2$ are
obtained from those for wave vector ${\bf q}_1$ by simply replacing 
${\bf q}_1$ by ${\bf q}_2$.

\begin{table} [h!]
\caption{\label{ACTIVE} The irrep labels are according to the tetragonal
structure. ``Pol" denotes polarization, where capitals (lower case)
refers to orthorhombic (tetragonal) coordinates, ``Int" denotes
intensity, and ``Modes" denotes the number of modes.  The columns labeled
${\bf q}_n$ apply when the wave vector ${\bf q}_n$ is selected.
$\sigma_1 = \sum_l \beta_{kl} \langle Q_{3,l}^-({\bf q})\rangle^2$,
$\sigma_2=\ \sum_{l,m} \gamma_{klm} \langle Q_{3,l}^-({\bf q}) \rangle^2
\langle Q_{2,m}^+({\bf q}) \rangle^2$,
and $\sigma_{3,k}=\sum_{lm} \langle Q_{2,l}^+({\bf q}) \rangle^2 
 \gamma_{klm} + \delta_{klm}
\langle Q_{3l}^-({\bf q}) \rangle^2]^2$, where $k$ is the mode index
and $\alpha$, $\beta$, and $\gamma$ do not depend on the OP's.}
\vspace{0.2 in}
\begin{tabular} {|| c | c c c | c c ||}
\hline \hline
Phase & Irrep & Int & Modes & \multicolumn{2} {|c|} {Pol} \\
\hline
& $\Gamma_5^-$ & $\alpha_k$ & 7 & \multicolumn{2} {|c||} {$x$} \\
\ \ I4/mmm\ \ & $\Gamma_5^-$ & $\alpha_k$ & 7 & \multicolumn{2} {|c||} {$y$} \\
& $\Gamma_5^-$ & $\alpha_k$ & 6 & \multicolumn{2} {|c||} {$z$} \\
\hline
&&&& \ \ ${\bf q}_1$\ \ &\ \ ${\bf q}_2$\ \ \\
\hline
& $X_1^+$ &$\sigma_{1k}$ & 5 & $Y$ & $Z$ \\
Cmcm & $X_2^+$ &$\sigma_{1k}$ & 2 & $Z$ & $Y$ \\
& $X_4^+$ &$\sigma_{1k}$ & 4 & $X$ & $X$ \\
\hline
& $\Gamma_1^+$ & $\sigma_{2k}$ & 4 & $Z$ & $Y$ \\
Cmc2$_1$ & $\Gamma_2^+$ &$\sigma_{2k}$ & 1 & $Y$ & $Z$ \\
& $\Gamma_5^+$ & $\sigma_{2k}$ & 5 & $X$ & $X$ \\
\hline
& $X_3^-$ & $\sigma_{3k}$ & 6 & $Z$ & $Y$ \\
Cmc2$_1$ & $X_4^-$ & $\sigma_{3k}$ & 6 & $Y$ & $Z$ \\
& $X_1^-$ & $\sigma_{3k}$ & 2 & $X$ & $X$ \\
\hline \hline
\end{tabular}

\end{table}

\section{Magnetoelastic Modes}

Here we investigate the mixing of dipole active phonons with
magnons for the RP system containing the magnetic ion Mn$^{4+}$.
The motivation for this investigation is that photon
absorption by magnons normally proceeds via a magnetic dipole
matrix element, whereas absorption by phonons proceeds via the very
much larger electric dipole matrix element.  In magnetic
ferroelectrics, magnons into which phonons are mixed have been
dubbed ``electromagnons."[\onlinecite{EM1}]
Such magnons not only have a much enhanced
absorption cross section, but, because their energy is often
much lower than the phonon energies, they can also lead to large anomalies
in the static dielectric constant.[\onlinecite{EM4}]
Potentially the same effect
is possible here if the magnons have a nontrivial coupling to phonons.

\begin{figure} [h!]
\begin{center}
\includegraphics[width=8.6cm]{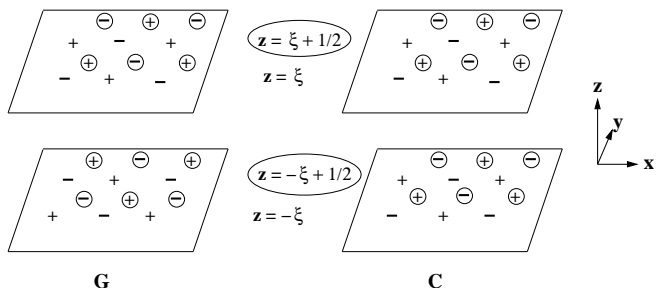}
\caption{\label{WK} Spin states of a bilayer of Mn ions at $z= \pm \xi$.
The plus and minus signs represent the signs of any component of spin.
Left: the ``G" configuration.  Right the ``C" configuration. 
The uncircled symbols are in the planes at $z=\pm \xi$ and the circled
symbols represent the spins in planes at $z=1/2 \pm \xi$ when
${\bf q}={\bf q}_1$.  For ${\bf q}={\bf q}_2$ the circled $+$ and $-$
signs are interchanged. The tetragonal axes are shown at right.}
\end{center}
\end{figure}

To start we review what is known about the magnetic structure of CMO which
appears below the magnetic ordering temperature of 115K.[\onlinecite{LOB}]
The OP's are[\onlinecite{LOB}] $G_X({\bf q}_1)$,  $C_Y({\bf q}_1)$, and $F_Z$,
or equivalently [\onlinecite{ABHCF}] $G_X({\bf q}_2)$,
$F_Y$, and $C_Z({\bf q}_2)$,
where $G$, $F$ and $C$ are the Wollan-Koehler magnetic structural
descriptors[\onlinecite{WK}]
for the magnetism of a single bilayer: F denotes a ferromagnetic
structure, G and C are the antiferromagnetic structures illustrated in Fig.
\ref{WK}.  The wave vector distinguishes between the two possible
stacking (of G and C) as one goes from one bilayer to the
next,[\onlinecite{ABHCF}] as shown in Fig. \ref{WK}. As mentioned
below Eq. (\ref{STRESSEQ}) one can select the wave vector by cooling
through $T_>$ in the presence of a shear stress.

The plan of our analysis is as follows.  We will construct the
most general trilinear coupling which involves two magnetic OP's
(in order to satisfy time-reversal invariance) and one phonon
operator whose wave vector is either at the zone center or at an $X$ point.
We will require invariance with respect to all the symmetries
of the tetragonal lattice.  (As a short-cut we fix the wave vector
to be ${\bf q}_1$.  The terms involving ${\bf q}_2$ can be determined
later by applying a four-fold rotation to those we find for ${\bf q}_1$.)
We then develop an effective bilinear coupling by replacing one of the
magnetic OP's by its value at the minimum of the free energy.  

Thus we analyze in detail the cubic interaction (involving the wave vector
${\bf q}_1$)
\begin{eqnarray}
V_{\rm ME}({\bf q}_1) &=& \sum_{\alpha \beta \gamma} [ A_{\alpha \beta \gamma}
C_\alpha ({\bf q}_1) M_\beta Q_\gamma^+ ({\bf q}_1) \nonumber \\ && \
+ B_{\alpha \beta \gamma}
C_\alpha ({\bf q}_1) N_\beta ({\bf q}_1) Q_\gamma^-  (0)
\nonumber \\ && \ + C_{\alpha \beta \gamma}
M_\alpha N_\beta ({\bf q}_1) Q_\gamma^-  ({\bf q}_1)] \nonumber \\
&\equiv& \sum_{k=1}^3 V_{\rm ME}^{(k)} ({\bf q}_1) \ ,
\label{SYMEQ} \end{eqnarray}
where ${\bf M}$ denotes ${\bf F}$, $\bf N ({\bf q})$ denotes 
${\bf G}({\bf q})$,
and we chose the arguments of the phonon operators so as to conserve
wave vector.  We also used the transformation properties of the magnetic
OP's under inversion (see Table \ref{SYM}) to infer the inversion signature
of the $Q$'s in Eq. (\ref{SYMEQ}).  

\begin{table} [h!]
\caption{\label{SYM} The symmetry for components $X$, $Y$, and $Z$ of the
magnetic OP's (which are pseudovectors).  Here $S_X=S_z$, $S_Y=S_x-S_y$, 
$S_Z=S_x+S_y$, where capitals refer to orthorhombic and lower case to
tetragonal coordinates. The results do not depend on the choice of wave vector.
In the text we use the standard notation in which the net magnetization,
here ${\bf F}$, is denoted ${\bf M}$ and the staggered magnetization,
here ${\bf G}({\bf q})$, is denoted ${\bf N}({\bf q})$.}
\vspace{0.2 in}
\begin{tabular} {||c c| c c c |c c c |c c c ||}
\hline \hline
Structure & \ \ ${\bf q}$\ \ & \multicolumn{3} {c|} {${\cal I}$} &
\multicolumn{3} {c|} {$m_z$} & \multicolumn{3} {c||} {$m_d$} \\
\hline
&& $Z$ & $Y$ & $X$ & $Z$ & $Y$ & $X$ & $Z$ & $Y$ & $X$ \\
\hline
F & $0$         & $+$ &$+$ &$+$ &  $-$ & $-$ & $+$ & $-$ & $+$ & $-$   \\
G & ${\bf q}$ & $-$ &$-$ &$-$ &  $+$ & $+$ & $-$ & $-$ & $+$ & $-$   \\
C & ${\bf q}$ & $+$ &$+$ &$+$ &  $-$ & $-$ & $+$ & $-$ & $+$ & $-$   \\
\hline \hline
\end{tabular}
\end{table}

\begin{table}
\caption{\label{MNQ} The symmetry of the magnetoeleastic trilinear
interaction for wave vector ${\bf q}_1$. All components are with
respect to orthorhombic coordinates. The variables $C$ and $N$ are
at wave vector ${\bf q}_1$, as are the irreps $X_k$.
Under $m_d$ ($m_z$) we give the effect of $m_d$ ($m_z$) on each of the
two magnetic OP's.  Then the irrep(s) is identified so as to make the
trilinear interaction invariant under $m_d$ and $m_z$ and in the next-to-last
column we identify the polarization $P$ of the absorption mode (if any).
Only irreps that are optically active are included in the analysis.
In the last column we give the order of magnitude of the term, assuming
the staggered moment to be unity.}
\vspace{0.2 in}
\begin{tabular} {||c c c | c c | c c | c | c || c ||}
\hline \hline
\multicolumn{3} {|| c |} {Interaction} & \multicolumn{2} {c|} {$m_d$} &
\multicolumn{2} {c|} {$m_z$} & irrep & $P$ & Int. \\
\hline
$M_Z$ & $N_Y$ & $Q^-({\bf q}_1)$ & $-$ & $+$ & $-$ & $+$ & 
$X_1^-$ & $X$ & $M^2\langle Q_2^+ ({\bf q}_1) \rangle^2$ \\
$M_Y$ & $N_X$ & $Q^-({\bf q}_1)$ & $+$ & $-$ & $-$ & $-$ & 
$X_3^-$ & $Z$ & $M^2 \langle Q_2^+({\bf q}_1)  \rangle^2$ \\
\hline
$C_X$ & $N_X$ & $Q^-(0)$ & $-$ & $-$ & $+$ & $-$ & 
$\Gamma_3^-$ & $X$ & $C^2$ \\
$C_Z$ & $N_X$ & $Q^-(0)$ & $-$ & $-$ & $-$ & $-$ & 
$\Gamma_5^-$ & $Z$ & $C^2$ \\
$C_Y$ & $N_Y$ & $Q^-(0)$ & $+$ & $+$ & $-$ & $+$ & 
$\Gamma_3^-$ & $X$ & $C^2$ \\
\hline
$M_X$ & $C_Y$ & $Q^+({\bf q}_1)$ & $-$ & $+$ & $+$ & $-$ & 
$X_4^+$ & $X$ & $C^2M^2\langle Q_3^-({\bf q}_1)  \rangle^2$ \\
$M_Y$ & $C_Y$ & $Q^+({\bf q}_1)$ & $+$ & $+$ & $-$ & $-$ & 
$X_1^+$ & $Y$ & $C^2M^2\langle Q_3^-({\bf q}_1)  \rangle^2$ \\
$M_Z$ & $C_Z$ & $Q^+({\bf q}_1)$ & $-$ & $-$ & $-$ & $-$ & 
$X_1^+$ & $Y$ & $C^2M^2 \langle Q_3^-({\bf q}_1)  \rangle^2$ \\
\hline \hline
\end{tabular}
\end{table}

To get the effective coupling between magnons and phonons we
only analyze special terms of Eq. (\ref{SYMEQ}).  We take the 
expectation value (that minimizes the free energy) of one of the
magnetic OP's and therefore arrive at a quadratic interaction between
the other (active) magnetic OP and a phonon OP.  To get a magnon-phonon
coupling out of this we select the component of the active magnetic OP
to be {\it transverse} to
its equilibrium value.  Since the nonzero equilibrium component of
${\bf N}({\bf q}_1)$ is its $X$ component, the transverse components of
${\bf N}({\bf q}_1)$ are its $Y$ and $Z$ components.  These are the
components of ${\bf N}({\bf q}_1)$ which, in the Holstein-Primakoff
formulation, create spin waves.  The operator $Q$ creates or destroys
a phonon.  These terms which mix magnons and phonons give
rise to ``electromagnons."[\onlinecite{EM1,EM2,EM3,EM4}]
We treat the other trilinear terms similarly, keeping
only those in which one magnetic OP is longitudinal and the other magnetic
OP is transverse.  Thereby we have the cases listed in Table \ref{MNQ}
for wave vector ${\bf q}_1$.  The results for wave vector ${\bf q}_2$ are
quite similar: the polarizations $Y$ and $Z$ are interchanged and
in the intensities ${\bf q}_1$ is replaced by ${\bf q}_2$, as in
Table \ref{ACTIVE}.
It is relevant to try to estimate which of these terms are the most
significant.  For that purpose we keep in mind the fact that
the OP's ${\bf M}$ and ${\bf C}({\bf q}_1)$ are roughly one tenth as large as
${\bf N}({\bf q}_1)$.[\onlinecite{LOB}]
In addition the optical activity of the
phonon depends on whether it is allowed in the tetragonal phase or
depends, for its activity, on the distortions of irreps $X_3^-$ or $X_2^+$
(in this connection we ignore $\Gamma_5^-$).  It seems likely, then,
that the largest magnon-phonon coupling will involve coupling the
transverse moment of $C({\bf q}_1)$ to either $X$ or $Z$ polarized
zone center phonons.  But for completeness, we list the other possibilities.
Table \ref{MNQ} only gives results for the case when wave vector ${\bf q}_1$
is selected.  Results for wave vector ${\bf q}_2$ can be obtained by
replacing $Z$ polarization by $Y$ and ${\bf q}_1$ by ${\bf q}_2$,
as is done in Table \ref{ACTIVE}.
 
These couplings can be detected either by anomalies in the static dielectric
constant when magnetic order appears,[\onlinecite{EM4}]
or by anomalous electric dipole intensity in magnons due to their coupling
with optical phonons.[\onlinecite{EM1,EM2,EM3,EM5}]

\section{CONCLUSIONS}

In this paper the effect on the optical properties of zone center phonons
of the sequential lowering of symmetry by structural
distortions from the high symmetry tetragonal (I4/mmm) phase has been studied.
In addition, the coupling of the magnons which appear at low
temperatures in the magnetically order phase of Ca$_3$Mn$_2$O$_7$ and
the phonons is explored to identify possible giant enhancements in
the magnon cross section in the electromagnetic spectrum.

Our specific results include

\noindent $\bullet 1$.
We have tabulated the zone center phonon modes with their polarizations
which appear in the absorption spectrum and in the Raman scattering spectrum
of the I4/mmm, Cmcm, and Cmc2$_1$ phases for Ca$_3$X$_2$O$_7$ systems, where
X is Mn or Ti.

\noindent $\bullet 2$.
For modes which become optically active by virtue of a structural distortion,
we indicate the dependence of their cross sections
on the newly emerging order parameters near the phase boundary
where the structural symmetry is lowered. We also discuss the
anomalies in the phonon frequencies as the phase boundaries are crossed.

\noindent $\bullet 3$.
We discuss the symmetry of the magnon phonon interaction which can lead
magnons being electric dipole active and hence having anomalously large
absorption cross section and leading to analogous large magnetic
contributions to the dielectric constant.

\begin{appendix}
\section{Intensity Calculation}
\label{APPA}
\subsection{Tetragonal Phase}
In this appendix I discuss in detail how the intensities of the modes
are calculated assuming the perturbative description relative to the
tetragonal phase is justified.  In the tetragonal phase we have the
optically active phonons of irreps $\Gamma_5^-$ and $\Gamma_3^-$,
governed by the Hamiltonian
\begin{eqnarray}
{\cal H}_{\rm TET} &=& \frac{1}{2} \sum_{k=1}^{n(\Gamma_5^-)}
\kappa_{\perp,k} (Y_k^2 + Z_k^2) + \frac{1}{2} 
\sum_{k=1}^{n(\Gamma_3^-)} \kappa_{\parallel,k} X_k^2 \ ,
\nonumber \end{eqnarray}
where it is conveneient to work in orthorhombic coordinates, so that
modes $X$, $Y$, and $Z$ are polarized along these respective orthorhombic
direction.  $X_k=Q_3^-(0)_k$, $Y_k=(Q_5^-(1)_k-Q_5^-(2)_k)/\sqrt 2$, and 
$Z_k=(Q_5^-(1)_k+Q_5^-(2)_k)/\sqrt 2$ are the order parameters for the modes
of the optically active $\Gamma_3^-$ and $\Gamma_5^-$ irreps.
To be specific: we number the modes in order of incresing energy.

The coupling to photons is via the dipole moment operator which we
write as
\begin{eqnarray}
{\bf p} &=& \sum_{k=1}^{n(\Gamma_5^-)} p_{\perp,k} [Y_k\hat Y +Z_k \hat Z] 
+ \sum_{k=1}^{n(\Gamma_3^-)} p_{\parallel,k} X_k \hat X\ ,
\nonumber \end{eqnarray}
where $\hat X$ is a unit vector in the orthorhombic $X$ direction and
similarly for $\hat Y$ and $\hat Z$.
So the absorption intensity in the tetragonal phase of the
$k$th doubly degenerate $Y_k$-$Z_k$ mode is $Ap_{\perp,k}^2$ and
that of the $k$th $X$ mode is $Ap_{\parallel,k}^2$,
where here and below we set $A=1$.  Since the dipole moment matrix
elements are subject to a first principles calculation, it makes
sense to give explicit results in terms of calculable quantities.

\subsection{Cmcm Phase}

The Hamiltonian that describes the mixing of new optically active
phonons in the Cmcm phase is
\begin{eqnarray}
{\cal H}_{\rm Cmcm} &=& \frac{1}{2} \sum_{k=1}^{n(X_1^+)}
\kappa^{(X_1^+)}_k Q_1^+({\bf q}_1)_k^2 
+\frac{1}{2} \sum_{k=1}^{n(X_2^+)} \kappa^{(X_2^+)}_k Q_2^+({\bf q}_1)_k^2 
\nonumber \\ && \
+\frac{1}{2} \sum_{k=1}^{n(X_4^+)} \kappa^{(X_4^+)}_k Q_4^+({\bf q}_1)_k^2 
\nonumber \\ && + \langle Q_3^-({\bf q}_1) \rangle
\Biggl[ \sum_{k=1}^{n[X_2^+]} \sum_{l=1}^{n[\Gamma_5^-]}
a^{(1)}_{k,l} Z_k Q_2^+ ({\bf q}_1)_l \nonumber \\ &&
+ \sum_{k=1}^{n[X_1^+]} \sum_{l=1}^{n[\Gamma_5^-]}
b^{(1)}_{k,l} Y_k Q_1^+ ({\bf q}_1)_l \nonumber \\ &&
\sum_{k=1}^{n[X_4^+]} \sum_{l=1}^{n[\Gamma_3^-]}
c^{(1)}_{k,l} Q_3^-(0)_k Q_4^+ ({\bf q}_1)_l \Biggr] \ .
\nonumber \end{eqnarray}
$Q_3^-({\bf q}_1)$ is replaced by $\langle Q_3^-({\bf q}_1)\rangle$
leading to a bilinear interaction which mixes new modes.

We will analyze the $Z$-component of the dipole moment operator.
The other components are treated analagously.  Due to the mixing
of modes, the bare mode operator $Z_k$ is related to the true
mode operators (indicated by tildes)
\begin{eqnarray}
Z_k &=& {\tilde Z}_k + \sum_l \frac{a^{(1)}_{kl} {\tilde Q}_2^+({\bf q}_1)_l}
{\kappa_l^{X_2^+} - \kappa_{\perp,k}} \langle Q_3^- ({\bf q}_1) \rangle \ .
\nonumber \end{eqnarray}
Thus the $Z$-component of the dipole moment operator which creates modes
${\tilde Q}_2^+({\bf q}_1)_l$ denoted $p_{Z,Q_2^+,l}$ is
\begin{eqnarray}
p_{Z,Q_2^+,l} &=& p_{\perp,k} \frac{
a^{(1)}_{kl} \langle Q_3^-({\bf q}_1)\rangle}
{ \kappa_l^{(X_2^+)} - \kappa_{\perp,k}} \langle Q_3^-({\bf q}_1) \rangle \ ,
\label{EQPP} \end{eqnarray}
and the intensity of this induced mode is $|p_{Z,Q_2^+,l}|^2.$
The quantities in Eq. (\ref{EQPP}) are amenable to a first principles
frozen-phonon calculation.
\end{appendix}

\noindent {\bf ACKNOWLEDGEMENTS.}
I would like to thank C. Fennie for a discussion of the properties of these
systems, for communicating the results of his first principle calculations,
and for pointing out many relevant references.
I am also grateful to B. Campbell and H. Stokes for discussions
concerning the group-subgroup relations and irreps for the star of
the wave vector.  I wish to thank M. Lobanov for discussions
concerning the magnetic structure of the Mn compound and for alerting me to
several references. I also thank G. Lawes for communicating the results of
the pyroelectric measurements.


\begin{thebibliography} {99}
\bibitem{FAM} 
It will be observed that the two families can be visualized as resulting from
the two equivalent orientations of the original sample which differ by a
four-fold rotation.  So the two families lead to two equivalent domain
structures.
\bibitem{FN4} 
In fact, Ref. \onlinecite{ABHPRB} proposed this mechanism to explain the
lack of such a correlation seen in experiments.  With only magnon-phonon
coupling the phonon energy shift in Eu$_{0.75}$Y$_{0.25}$MnO$_3$
would be expected to be approximately 20 cm$^{-1}$ whereas it is observed
to be only 2 cm$^{-1}$.[\onlinecite{EM5}]
\end{thebibliography}
\end{document}